\documentclass[]{emulateapj}
\slugcomment{ApJS in press }

\newcommand{\teff}{\ensuremath{T_{\rm eff}}}
\newcommand{\logg}{\ensuremath{\log{g}}}
\newcommand{\vsini}{\ensuremath{V \sin i}}
\newcommand{\feh}{\rm [Fe/H]}

\newcommand{\rsun}{\ensuremath{R_\sun}}
\newcommand{\msun}{\ensuremath{M_\sun}}
\newcommand{\lsun}{\ensuremath{L_\sun}}

\newcommand{\rstar}{\ensuremath{R_\star}}
\newcommand{\mstar}{\ensuremath{M_\star}}

\newcommand{\loglstar}{\ensuremath{\log{L_\star}}}
\newcommand{\rhostar}{\ensuremath{\rho_\star}}

\newcommand{\rearth}{\ensuremath{R_\oplus}}
\newcommand{\mearth}{\ensuremath{M_\oplus}}

\def\note #1]{{\bf #1]}}

\newcommand{\starname}{Kepler-18}
\newcommand{\planetb}{Kepler-18b}  
\newcommand{\planetc}{Kepler-18c}  
\newcommand{\planetd}{Kepler-18d}  
\newcommand{\kicid}{KIC~8644288}
\newcommand{\tmid}{2MASS J19521906\linebreak[0]+4444467}
\newcommand{\wspitzer}{\emph{Warm-Spitzer}}
\newcommand{\spitzer}{\emph{Spitzer}}

\newcommand{\kepmag}{13.549}

\newcommand{\reconteff}{\ensuremath{5250 \pm 125}}
\newcommand{\reconlogg}{\ensuremath{4.0 \pm 0.25}}
\newcommand{\reconvsini}{\ensuremath{2 \pm 2}}
\newcommand{\SMEteff}{\ensuremath{5383 \pm 44}}
\newcommand{\SMEfeh}{\ensuremath{+0.20 \pm 0.04}}
\newcommand{\SMElogg}{\ensuremath{4.41 \pm 0.10}}
\newcommand{\SMEvsini}{\ensuremath{0.4 \pm 0.5}}
\newcommand{\teffshort}{\ensuremath{5345}}
\newcommand{\MOOGteff}{\ensuremath{5345 \pm 100}}
\newcommand{\MOOGfeh}{\ensuremath{+0.19 \pm 0.06}}
\newcommand{\MOOGlogg}{\ensuremath{4.31 \pm 0.12}}
\newcommand{\MOOGvsini}{\ensuremath{< 4}}
\newcommand{\epochb}{\ensuremath{2454966.5068 \pm 0.0021}}
\newcommand{\periodb}{\ensuremath{3.504725 \pm 0.000028}}
\newcommand{\impactb}{\ensuremath{0.771 \pm 0.025}}
\newcommand{\scaledPlanetRadiusb}{\ensuremath{0.01656 \pm 0.00032}}
\newcommand{\MCMCmplanetb}{\ensuremath{13.4 \pm 5.8}}
\newcommand{\TTVmplanetb}{\ensuremath{6.9 \pm 3.4}}
\newcommand{\rplanetb}{\ensuremath{2.00 \pm 0.10}}

\newcommand{\TTVrhoplanetb}{\ensuremath{4.9 \pm 2.4}}

\newcommand{\TTVepochc}{\ensuremath{2455167.0883 \pm 0.0023}}

\newcommand{\TTVperiodc}{\ensuremath{7.64159 \pm 0.00003}}
\newcommand{\impactc}{\ensuremath{0.593 \pm 0.050}}
\newcommand{\scaledPlanetRadiusc}{\ensuremath{0.04549 \pm 0.00055}}
\newcommand{\MCMCmplanetc}{\ensuremath{16.9 \pm 6.1}}
\newcommand{\TTVmplanetc}{\ensuremath{17.3 \pm 1.9}}
\newcommand{\rplanetc}{\ensuremath{5.49 \pm 0.26}}

\newcommand{\TTVrhoplanetc}{\ensuremath{0.59 \pm 0.07}}

\newcommand{\TTVepochd}{\ensuremath{2455169.1776 \pm 0.0013}}

\newcommand{\TTVperiodd}{\ensuremath{14.85888 \pm 0.00004}}
\newcommand{\impactd}{\ensuremath{0.767 \pm 0.024}}
\newcommand{\scaledPlanetRadiusd}{\ensuremath{0.05782 \pm 0.00069}}
\newcommand{\MCMCmplanetd}{\ensuremath{29.9  \pm 8.8}}
\newcommand{\TTVmplanetd}{\ensuremath{16.4 \pm 1.4}}
\newcommand{\rplanetd}{\ensuremath{6.98 \pm 0.33}}

\newcommand{\TTVrhoplanetd}{\ensuremath{0.27  \pm 0.03}}
\newcommand{\kepler}{\emph{Kepler}}
\newcommand{\blender}{{\tt BLENDER}}
\newcommand{\koiname}{K00137}
\newcommand{\koione}{K00137.01}
\newcommand{\koitwo}{K00137.02}
\newcommand{\koithree}{K00137.03}

\begin{document}

\shortauthors{Cochran et al.}
\shorttitle{\planetb,c,d}


\bibliographystyle{apj}

\title{{\planetb, \lowercase{c, and d}}:
A System Of Three Planets Confirmed by Transit Timing Variations,
Lightcurve Validation, {\wspitzer} Photometry and Radial Velocity Measurements}

\author{
William D. Cochran\altaffilmark{1},
Daniel C. Fabrycky\altaffilmark{2},
Guillermo Torres\altaffilmark{3},
Fran\c{c}ois Fressin\altaffilmark{3},
Jean-Michel D{\'e}sert\altaffilmark{3},
Darin Ragozzine\altaffilmark{3},
Dimitar Sasselov\altaffilmark{3},
Jonathan J. Fortney\altaffilmark{2},
Jason F. Rowe\altaffilmark{4},
Erik J. Brugamyer\altaffilmark{5},
Stephen T. Bryson\altaffilmark{4},
Joshua A. Carter\altaffilmark{3},
David R. Ciardi\altaffilmark{6},
Steve B. Howell\altaffilmark{4},
Jason H. Steffen\altaffilmark{7},
William. J. Borucki\altaffilmark{4},
David G. Koch\altaffilmark{4},
Joshua N. Winn \altaffilmark{8},
William F. Welsh\altaffilmark{9},
Kamal Uddin\altaffilmark{10,4},
Peter Tenenbaum\altaffilmark{15,4},
M. Still\altaffilmark{11,4},
Sara Seager\altaffilmark{8},
Samuel N. Quinn\altaffilmark{3},
F. Mullally\altaffilmark{15,4},
Neil Miller\altaffilmark{2},
Geoffrey W. Marcy\altaffilmark{12},
Phillip J. MacQueen\altaffilmark{1},
Philip Lucas\altaffilmark{13},
Jack J. Lissauer\altaffilmark{4},
David W. Latham\altaffilmark{3},
Heather Knutson\altaffilmark{12},
K. Kinemuchi\altaffilmark{11,4},
John A. Johnson\altaffilmark{14},
Jon M. Jenkins\altaffilmark{15,4},
Howard Isaacson\altaffilmark{12},
Andrew Howard\altaffilmark{12},
Elliott Horch\altaffilmark{16},
Matthew J. Holman\altaffilmark{3},
Christopher E. Henze\altaffilmark{4},
Michael R. Haas\altaffilmark{4},
Ronald L. Gilliland\altaffilmark{17},
Thomas N. Gautier III\altaffilmark{18},
Eric B. Ford\altaffilmark{19},
Debra A. Fischer\altaffilmark{20},
Mark Everett\altaffilmark{21},
Michael Endl\altaffilmark{1},
Brice-Oliver Demory\altaffilmark{8},
Drake Deming\altaffilmark{22},
David Charbonneau\altaffilmark{3},
Douglas Caldwell\altaffilmark{15,4},
Lars Buchhave\altaffilmark{23,24},
Timothy M. Brown\altaffilmark{25},
and
Natalie Batalha\altaffilmark{26}
}
\altaffiltext{1}{McDonald Observatory, The University of Texas, Austin,
	TX 78712}
\altaffiltext{2}{Department of Astronomy and Astrophysics, University of
	California, Santa Cruz, CA 95064, USA}
\altaffiltext{3}{Harvard-Smithsonian Center for Astrophysics, 60 Garden
	Street, Cambridge, MA 02138; jdesert@cfa.harvard.edu}
\altaffiltext{4}{NASA Ames Research Center, Moffett Field, CA 94035}
\altaffiltext{5}{Department of Astronomy, The University of Texas, Austin,
	TX 78712}
\altaffiltext{6}{NASA Exoplanet Science Institute/CalTech, Pasadena, CA 91125 USA}
\altaffiltext{7}{Fermilab Center for Particle Astrophysics, P.O. Box 500,
	MS 127, Batavia, IL 60510}
\altaffiltext{8}{Massachusetts Institute of Technology, Cambridge, MA 02139 USA}
\altaffiltext{9}{San Diego State Univ., San Diego, CA 92182 USA}
\altaffiltext{10}{Orbital Sciences Corporation}
\altaffiltext{11}{Bay Area Environmental Research Inst.}
\altaffiltext{12}{Department of Astronomy, University of California,
	Berkeley, CA 94720-3411, USA}
\altaffiltext{13}{Centre for Astrophysics Research, University of
	Hertfordshire, College Lane, Hatfield, AL10 9AB, England}
\altaffiltext{14}{California Institute of Technology, Pasadena, CA 91109}
\altaffiltext{15}{SETI Institute, Mountain View, CA, 94043, USA}
\altaffiltext{16}{Southern Connecticut State University, New Haven, CT, 06515}
\altaffiltext{17}{Space Telescope Science Institute, Baltimore, MD, 21218, USA}
\altaffiltext{18}{Jet Propulsion Laboratory, Calif. Institute of Technology,
	Pasadena, CA, 91109, USA}
\altaffiltext{19}{University of Florida, Gainesville, FL 32611}
\altaffiltext{20}{Yale University, New Haven, CT 06510}
\altaffiltext{21}{NOAO, 950 N. Cherry Ave., Tucson, AZ 85719}
\altaffiltext{22}{Solar System Exploration Division, NASA Goddard Space
	Flight Center, Greenbelt, MD 20771, USA}
\altaffiltext{23}{Niels Bohr Institute, University of Copenhagen, DK-2100
	Copenhagen, Denmark}
\altaffiltext{24}{Centre for Star and Planet Formation, Natural History
	Museum of Denmark, University of Copenhagen, DK-1350 Copenhagen,
	Denmark}
\altaffiltext{25}{Las Cumbres Observatory Global Telescope, Goleta, CA 93117}
\altaffiltext{26}{San Jose State University, San Jose, CA, 95192, USA}
\altaffiltext{$\dagger$}{Based in part on observations obtained at the W.~M.~Keck Observatory, which is operated by the University of California and the California Institute of Technology.}
\altaffiltext{*}{To whom correspondence should be addressed.  E-mail:
wdc@astro.as.utexas.edu}

\begin{abstract}

We report the detection of three transiting planets around a Sunlike
star, which we designate {\starname}. The transit signals were detected in
photometric data from the {\kepler} satellite, and were confirmed to
arise from planets using a combination of large transit-timing
variations, radial-velocity variations, {\wspitzer} observations, and
statistical analysis of false-positive probabilities. The {\starname}
star has a mass of 0.97\msun, radius 1.1\rsun, effective temperature
{\teffshort}\,K, and iron abundance [Fe/H]= +0.19. 
The planets have orbital periods of approximately 3.5, 7.6 and 14.9 days.
The innermost planet ``b'' is a ``super-Earth'' with mass
\TTVmplanetb\mearth, radius \rplanetb\rearth, and mean density
{\TTVrhoplanetb}\,g cm$^{−3}$.  The two outer planets
``c'' and ``d'' are both low-density Neptune-mass planets.
{\planetc} has a mass of \TTVmplanetc\mearth, radius \rplanetc\rearth,
and mean density {\TTVrhoplanetc}\,g cm$^{−3}$,
while {\planetd} has a mass of \TTVmplanetd\mearth, radius \rplanetd\rearth, 
and mean density {\TTVrhoplanetd}\,g cm$^{−3}$.
{\planetc} and {\planetd} have orbital periods near a 2:1 mean-motion
resonance, leading to large and readily detected transit timing variations.

\end{abstract}



\section{Introduction}\label{sec:intro}

{\kepler} is a NASA Mission designed to detect the transits of exoplanets
across the disks of their stars.  The ultimate mission goal is to detect the
transits of potentially habitable Earth-size planets.  To achieve this goal
requires a telescope in a very stable space environment with a large
(0.95-meter) effective aperture monitoring the brightness of about 150,000
stars simultaneously and continuously for over three years.
The {\kepler} Mission design and performance are summarized by \citet{BoKoBa10}
and by \citet{KoBoBa10}, and a discussion of the commissioning and first
quarter data are given by \citet{BoKoBa11a}.
\citet{BoKoBa11b} reported 1235 planet candidates that were discovered during
the first four months of the Mission.  \citet{BaBoKo10} discuss the
selection and characteristics of the {\kepler} target stars.

The first 5 planets discovered by the {\kepler} mission (Kepler\,4-8)
were reported in January 2010
\citep{BoKoBr10,KoBoRo10,DuBoKo10,LaBoKo10,JeBoKo10}.
{\kepler} has detected an abundance of multi-planet systems.
\citet{BoKoBa11b} reported a total of 170 candidate multi-planet systems
among the 997 planet candidate host stars from the February 2011 data release.
\citet{StBaBo10} presented five of these systems in detail.
The Kepler-9b and c system \citep{HoFaRa10} was the first transiting
multi-planet system confirmed by transit timing variations.
Kepler-10 \citep{BaBoBr11} was the first rocky planet found by \kepler.
Kepler-11 \citep{LiFaFo11} is a transiting system of six planets.

Not all {\kepler} planets can be directly confirmed by supporting reflex
radial velocity measurements of the parent star, or by detection and modeling
of transit timing variations.  Instead, some planets must be validated
by analyzing all possible possible astrophysical false-positive scenarios
and comparing their {\itshape a priori} likelihood to that of a planet.
The {\kepler} project has been able to
utilize the {\blender} technique developed by \citet{ToKoSa04} to
validate a third planet in the Kepler-9 system \citep{ToFrBa11},
a second planet in the Kepler-10 \citep{FrToDe11} and the outer
planet in the Kepler-11 system \citep{LiFaFo11}.

Here we present the {\starname}  system, containing two Neptune-mass
transiting planets near a 2:1 mean motion resonance which show significant
gravitational interactions which are observed via measurements of transit
timing variations (TTVs), as well as a small, inner super-Earth size
transiting planet.  This system is remarkably similar to the Kepler-9
system in its overall architecture.

\section{{\kepler} Photometry}\label{sec:photometry}
The {\kepler} spacecraft carries a photometer with a wide-field
($\sim 115 \deg^2$) Schmidt camera of 0.95-m effective aperture.
The spacecraft was launched in March 2009, and is now in an
Earth-trailing heliocentric orbit which allows nearly continuous
photometric coverage of its field-of-view in Cygnus and Lyra.
\citet{CaKoVC10} discuss the early instrumental performance of
the {\kepler} photometer system.
The primary data for detection of transiting planets are the
Long Cadence (LC) ``Pre-search Data Conditioned'' (PDC) time series data,
in which 270 consecutive CCD readouts are binned, giving an effective
sampling interval of 29.4244~minutes \citep{JeCaCh10a}.
A small selected subset of {\kepler} targets is sampled at the Short Cadence
(SC) rate of 9 consecutive reads for a sampling interval of 58.85 seconds
\citep{GiJeBo10}.  Thus, one LC sample is the sum of 30 SC samples.
The data from the spacecraft are processed through the {\kepler} Science
Operations Center pipeline \citep{JeCaCh10b} to perform standard CCD
processing and to remove instrumental artifacts.  The LC PDC time series data
are searched for possible planetary transits using a wavelet-based
adaptive matched filter \citep{JeCChMC10}.  Possible planetary transit
events with amplitude greater than $7.1 \sigma$ are flagged and are then
subjected to intensive validation efforts using the {\kepler} data
\citep{BaRoGi10,WuTwTe10}.  Objects that pass this level of vetting are
designated as a ``Kepler Object of Interest'' (KOI) and are sent
to the Follow-up Observing Program (FOP) for further study.

\subsection{Light Curves and Data Validation}\label{sec:lightCurves}
One of the objects identified with possible transiting planets is 
the Kp {\kepmag} magnitude (where Kp is the magnitude in 
the {\kepler} passband) star \kicid\ (\tmid, \koiname).
After a possible transiting planet has been detected,
the {\kepler} data are subjected to a set of statistical tests to search
for possible astrophysical false-positive origin of the observed signal.
These data validation tests for the first five {\kepler} planet discoveries
are described by \citet{BaRoGi10}.   Additional tests, including
measurement of the image centroid motion during a transit \citep{WuTwTe10}
all gave a high probability that the signals seen were real.
The application of these techniques to Kepler-10b is described
in detail by \citet{BaBoBr11}.

\begin{figure*}
\includegraphics[angle=-90,width=\textwidth]{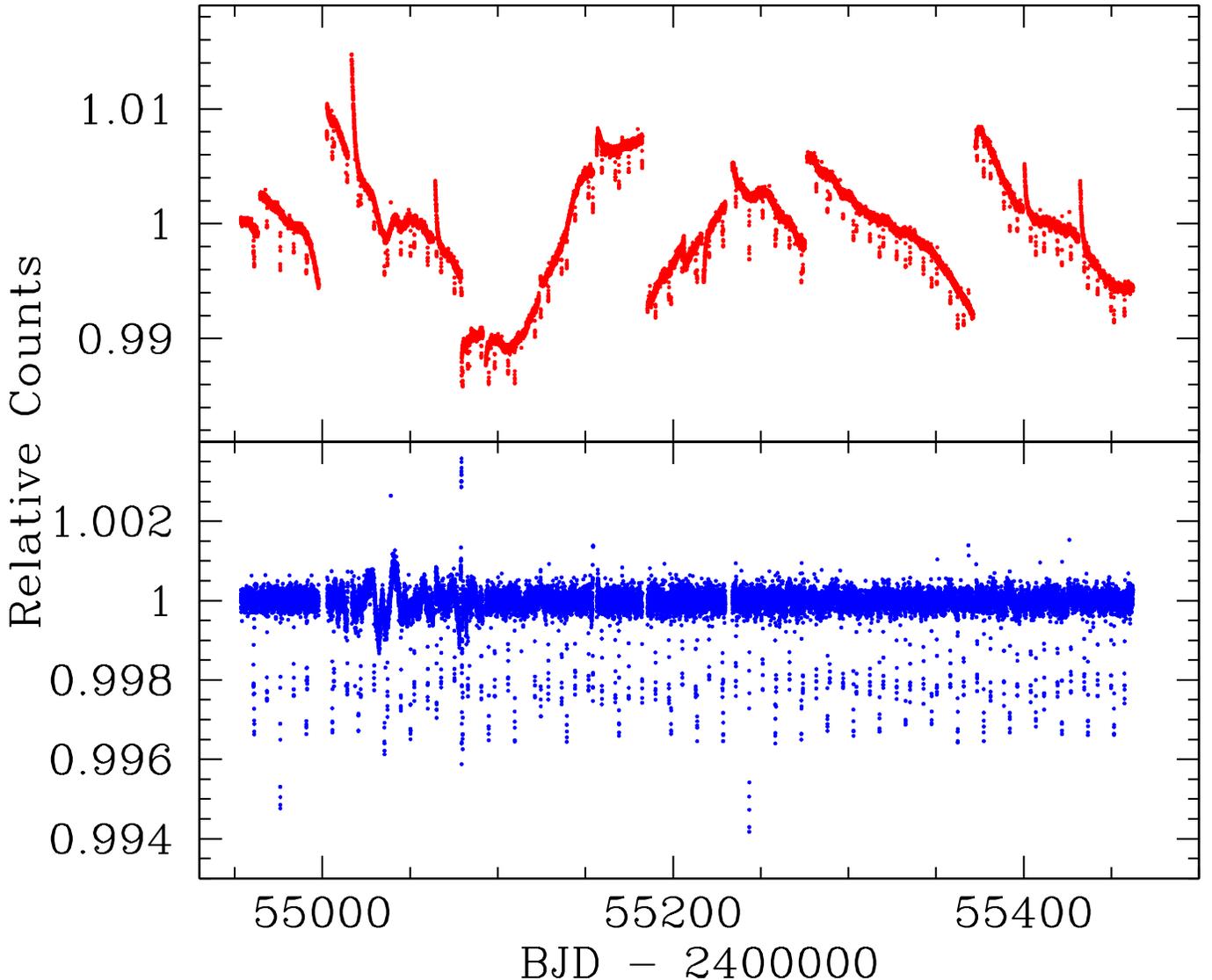}
\caption{{\kepler}  lightcurves for \koiname.
The upper panel shows the normalized raw Photometric Analysis
(PA) lightcurves for quarters 0 through 6.
The various discontinuities are due to effects such as spacecraft
safe-mode events or loss of fine pointing.
Each quarter put the star on a different detector, which
accounts for the change in overall sensitivity.
The long-term drifts in each quarter are temperature related.
The lower panel shows the normalized corrected Pre-search Data Conditioned
(PDC) lightcurve.  Most of the spacecraft-related
variability of the PA lightcurve has been removed.
\label{fig:lightcurve}}
\end{figure*}

Two separate transiting objects were immediately obvious in the LC data.
{\koione}  has a transit ephemeris of
$\rm{T_0[BJD]} = (\TTVepochc) + N * (\TTVperiodc)$\,days and a transit depth of
$2287 \pm 9$\,ppm.
(All transit times and ephemerides in this paper are based on UTC.)
{\koitwo} has an ephemeris of
$\rm{T_0[BJD]} = (\TTVepochd) + N * (\TTVperiodd)$\,days
and a transit depth of $3265 \pm 12$\,ppm.
It was noted that the orbits of these two objects were very near a period
ratio of 2:1.
After filtering these transits of {\koione} and {\koitwo} from the lightcurve,
we searched again for transiting objects, and found a third planet
candidate in the system, {\koithree}, which has a shorter orbital period than
the other two transiting planets.
{\koithree} has the ephemeris
$\rm{T_0[BJD]} = (\epochb) + N * (\periodb)$\,days,
and a transit depth of only $254 \pm 8$\,ppm.
The light curve for {\koiname} is shown in Figure~\ref{fig:lightcurve}.
The upper panel shows the raw uncorrected ``Photometric Analysis'' (PA)
lightcurves that come out of the data processing pipeline,
and the lower panel shows the corrected PDC lightcurves.
The two transit events that look significantly deeper than the others,
near BJD\,2454976.0 and BJD\,2455243.5 are simultaneous transits of {\koione}
and {\koitwo}.

%
%

\begin{figure}
\plotone{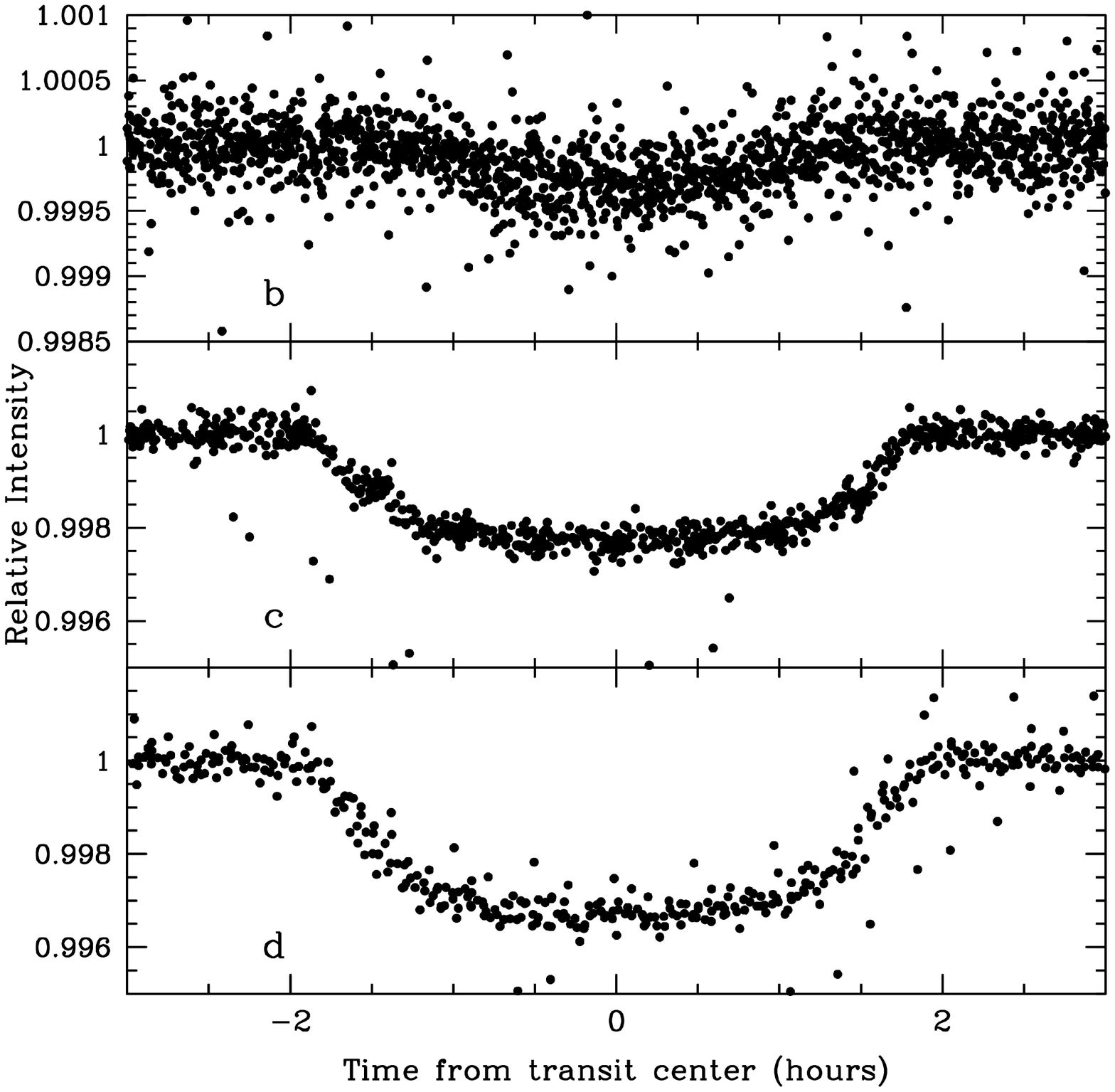}
\caption{Folded Lightcurves for \koiname.
The top row is {\planetb} ({\koithree}),
the middle row is {\planetc} (\koione)
and the bottom row is {\planetd} (\koitwo).
The lightcurves are folded on
the mean period listed in Table~\ref{tab:SystemParameters}.
\label{fig:foldedLCs}}
\end{figure}

The folded lightcurves for each of the three candidate planets are shown
in Figure~\ref{fig:foldedLCs}.  This Figure shows
the lightcurves folded on the ephemeris given above for each KOI.
The significant width of the ingress and egress for {\planetc} and {\planetd}
are a result of the transit time variations discussed in Section~\ref{sec:TTV}.

%
%

\section{Follow-Up Observations}\label{sec:fop}
After the possible transiting planets were found, and the KOIs passed the
Data Validation tests for false positive signals, {\koiname} was sent on to
the {\kepler}  Follow-up Observing Program (FOP) for ground-based telescopic
observations designed either to find any additional indication that these
KOIs might be an astrophysical false-positive signal,
or to verify the planetary nature of the transit events.

\subsection{Reconnaissance Spectroscopy}\label{sec:recon}
The first FOP step is to obtain high spectral resolution,
low S/N spectroscopy in order to verify the {\kepler}  Input Catalog
stellar classification, and
to search for evidence of stellar multiplicity in the spectrum.
Spectra from the Hamilton echelle spectrograph on the Lick Observatory
3-m Shane Telescope were obtained on the nights of 8 and 9 August 2009,
and on 1 September 2009 UT.  These spectra showed no convincing evidence
for radial velocity variability at the 0.5\,km\,s$^{-1}$ level,
and no hints of any contaminating spectra.
These spectra were cross-correlated against a library of synthetic stellar
spectra as described by \citet{BaBoBr11},
in order to derive basic stellar parameters to compare with the
Kepler Input Catalog (KIC) values.
These spectra yielded $\teff = \reconteff$K, $\logg = \reconlogg$, and
$\vsini = \reconvsini$\,km\,s$^{-1}$.
The height of the cross-correlation peaks ranged from 0.82 to 0.89,
indicating a very good match with the library spectra.

\subsection{High Spatial-Resolution Imaging}\label{sec:imaging}

High resolution imaging of the surroundings of a KOI is an important
step to identify possible sources of false-positive signals.
We need to ensure that the detected transit signal is indeed originating
on the selected target star, and not on a background star that was
unresolved in the original KIC imaging of the {\kepler} field.

A seeing limited image was obtained at Lick Observatory's 1-m Nickel
telescope using the Direct Imaging Camera.  This image was a single
one-minute exposure taken in the I-band, with seeing of approximately
1.5\arcsec.  The only nearby star is a faint object (approximately
5\,magnitudes fainter than \koiname) about {5.5\arcsec} to the north.
A J-band image of the $1\arcmin \times 1\arcmin$ field of view around
\koiname was taken as part of a complete J-band survey of the {\kepler}
field of view using the wide field camera (WFCAM) on the United Kingdom
Infrared Telescope (UKIRT).
These images have a typical spatial resolution of 0.8-0.9{\arcsec}
and an image depth of J = 19.6 (Vega system).
This image also shows only a faint source about 5.5{\arcsec} to the north.

%
%

\begin{figure}[!ht]
\plotone{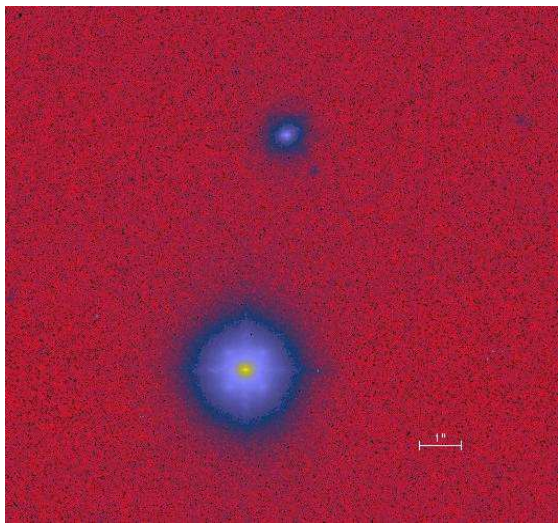}
\caption{High spatial resolution A/O image of {\koiname} taken with the
Palomar Observatory 5m adaptive optics system.  The nearest contaminating
source is about 5.5{\arcsec} to the north.
\label{fig:images}}
\end{figure}

We perform speckle observations at the WIYN 3.5-m telescope,
using the Differential Speckle Survey Instrument (DSSI)
\citep{HoGoSh11,HoEvSh11}.  The DSSI provides simultaneous
observations in two filters by employing a dichroic beam splitter and
two identical EMCCDs as the imagers. We generally observe simultaneously in
V and R bandpasses, where V has a central wavelength of 5620{\AA},
and R has a central wavelength of 6920{\AA}, and each filter has a
FWHM of 400{\AA}. 
The speckle observations of {\koiname} were obtained on 22 June 2010 UT
and consisted of three sets of 1000, 80\,msec individual speckle images.
Along with a nearly identical V-band reconstructed image, the speckle
results reveal no companion star near {\koiname} within the annulus from
0.05{\arcsec} to 1.8{\arcsec} to a limit (5$\sigma$) of 4.2 magnitudes fainter
in both V and R than the Kp {\kepmag} target star.

Near-infrared adaptive optics imaging of {\koiname} was obtained on the night
of 08~September 2009 UT with the Palomar Hale 5-m telescope and the PHARO
near-infrared camera \citep{HaBrPi01} behind the Palomar adaptive optics
system \citep{TrDeBr00}.  PHARO, a $1024\times20124$ HgCdTe infrared array,
was utilized in 25.1 mas/pixel mode yielding a field of view of
25{\arcsec}.  Observations were performed in $J$ filter.
The data were collected in a standard 5-point quincunx dither
pattern of 5{\arcsec} steps interlaced with an off-source (60{\arcsec} East)
sky dither pattern. Data were taken with integration times per frame
of 30\,sec with a total on-source integration time of 7.5\,minutes.
The adaptive optics system guided on the primary target itself and
produced  a central  core width of $FWHM = 0.075\arcsec$.
Figure~\ref{fig:images} shows this Palomar image of {\koiname}.
There are two additional sources within 15{\arcsec} of the primary
target. The first source, located 5.6{\arcsec} to the north of {\koiname},
is 4 magnitudes fainter at J; the second, located 15.0{\arcsec} to the
southeast, is 5.8 magnitudes fainter.  To produce the observed
transit depths in the blended {\kepler} photometry, the $\delta_J = 4$ mag
star would need to have eclipses that are 0.09 and 0.5 magnitudes deep, and
the $\delta J = 5.8$ mag star would need to have eclipses that are 0.15 and
1.3 magnitudes deep.  Such deep eclipses from stars separated by more than
4{\arcsec} ($>1$ {\kepler} pixel) would easily be detected in the centroid
motion analysis, but no centroid motion was detected between in and out of
transit, indicating that these two stars were not responsible for the
observed events.  Six additional sources were detected at $J$ within
15{\arcsec} of the primary target.  But all of these sources are
$\delta J>8$ magnitudes fainter than {\koiname} and could not produce the
observed  transit events (i.e., even if the stars dimmed by 100\%, the
resulting transit in the blended photometry would not be deep enough to
match the observed  transit depths).

Source detection completeness was evaluated by randomly inserted fake
sources of various magnitudes in steps of 0.5 mag and at varying distances
in steps of 1.0 FWHM from the primary target. Identification of sources was
performed both automatically with the IDL version of DAOPhot and by eye.
Magnitude detection limits were set when a source was not detected by the
automated FIND routine or was not detected by eye.  Within a distance of
$1-2$ FWHM, the automated finding routine often failed even though the eye
could discern two sources, but beyond that distance the two methods agreed
well.  A summary of the detection efficiency as a function of distance from
the primary star is given Table~\ref{tab:paloAOlimits}.

%
%
 
\begin{deluxetable}{cccc}
\tablenum{1}
\tablewidth{0pt}
\tablecaption{Palomar AO source sensitivity as a function of
distance from the primary target at $J$.
\label{tab:paloAOlimits}}
\tablehead{\colhead{Distance} &  \colhead{Distance} & \colhead{$\Delta$J} &
\colhead{J} \\
\colhead{(FWHM)} & \colhead{$(\arcsec)$} & \colhead{(mag)} & \colhead{(mag)}}
\startdata
1  &  0.075  &  1.5   & 13.7 \\
2  &  0.150  &  2.5   & 14.7 \\
3  &  0.225  &  3.5   & 15.7 \\
4  &  0.300  &  4.0   & 16.2 \\
5  &  0.375  &  4.5   & 16.7 \\
6  &  0.450  &  5.0   & 17.2 \\
7  &  0.525  &  6.0   & 18.2 \\
8  &  0.600  &  7.0   & 19.2 \\
9  &  0.675  &  7.5   & 19.7 \\
40 &  3.000  &  8.5   & 20.7
\enddata
\end{deluxetable}

A major source of false positive planet indication in the {\kepler} data
is background eclipsing binary (BGEB) stars within the photometric
aperture of {\starname}, which, when diluted by {\starname} itself,
can produce a planetary-size transit signal. 
We perform a direct measurement of the source location via difference images.
Difference image analysis takes the difference between average in-transit
pixel images and average out-of-transit images. 
Barring pixel-level systematics, the pixels with the highest flux in the 
difference image will form a star image at the location of the transiting
object, with amplitude equal to the depth of the transit.
Performing a fit of the the Kepler pixel response function (PRF)
\citep{BrTeJe10} to both the difference and out-of-transit images
quantifies the offset of the transit source from {\starname}.
Difference image analysis is vulnerable to various systematics due to
crowding and PRF errors which will bias the result \citep{BrTeJe10}. 
These types of biases will vary from quarter to quarter.
We ameliorate these biases by computing the uncertainty-weighted
robust average of the source locations over available quarters.  
Table \ref{tab:planet_offsets} gives the offsets of the transit
signal source from {\starname} averaged over
quarters 1 through 8 for all three planet candidates.  We see that 
the average offsets are within 1 sigma of {\starname},
with {\planetb} being just over $2 \sigma$.  
From all of these lines of evidence, we conclude that there are no other
objects withing 4\,mags near {\koiname} from 0.05 arcsec out to 15 arcsec.

%
%

\begin{deluxetable}{lcc}
\tablenum{2}
\tablewidth{0pt}
\tablecaption{Mean Pixel Response Function Fit Source Offsets \label{tab:planet_offsets}}
\tablehead{\colhead{Planet} & \colhead{Distance (arcsec)} & \colhead{$\sigma$}}
\startdata
\planetb & $0.536 \pm 0.245$ & 2.19 \\
\planetc & $0.070 \pm 0.107$ & 0.66 \\
\planetd & $0.067 \pm 0.103$ & 0.65 \\
\enddata
\end{deluxetable}

\subsection{Precise Doppler Measurements of \starname}\label{sec:rv}
\begin{figure}
\epsscale{1.2}
\plotone{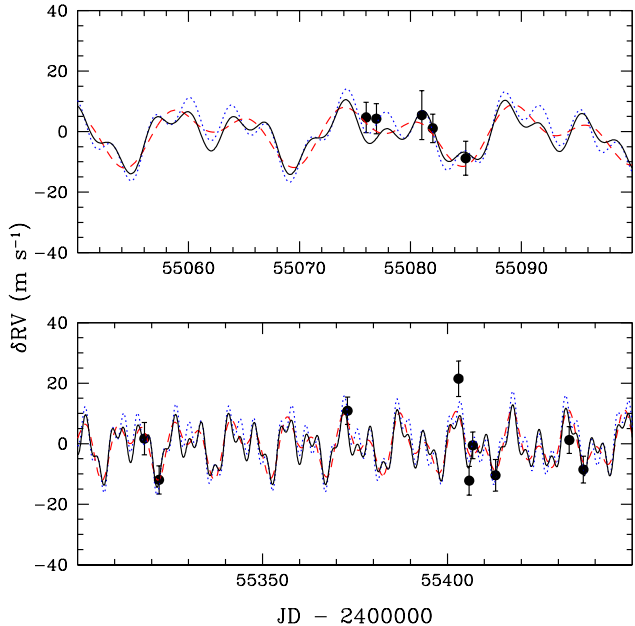}
\caption{Observed precise relative radial velocity measurements from
the Keck HIRES spectrograph.  The solid black line gives the relative 
radial velocity variations of {\starname} computed from the full
multi-body dynamical model discussed in Section~\ref{sec:TTV}.  
The dotted (blue) line is the model fit from the
combined MCMC solution to the {\kepler} lightcurve and these velocity
data presented in Section~\ref{sec:LCRV}.
A Keplerian fit from the radial velocity data alone, without the lightcurve
constraints, is indistinguishable from this curve.
The dashed (red) line is a two-planet Keplerian fit to the RV data accounting
for {\planetc} (\koione) and {\planetd} (\koitwo) only.
\label{fig:RVs}}
\end{figure}

After completion of the reconnaissance spectroscopy and high resolution
imaging, {\koiname} showed no evidence that might refute the planetary nature
of the transit event in the {\kepler} lightcurve and the target was approved
for high precision radial velocity observations.
We obtained 14 relative velocities with the Keck\,1 HIRES spectrometer
\citep{VoAlBi94} between 2009 September 1 and 2010 August 28 UT.
We used the same spectrograph configuration as is normally used for
precise Doppler measurements of solar type stars
\citep[cf.][]{MaBuVo08,CoHaPa02}.
Decker B5 ($0.87\arcsec  \times 3.5\arcsec$) was used for the first
five RV observations taken in 2009, and decker C2
($0.87\arcsec \times 14\arcsec$) was used for all of the later spectra.
Exposures taken with the B5 decker entrance to the HIRES spectrometer
suffer RV errors of up to $\pm15$\,m\,s$^{-1}$  when the moon is full
and the relative Doppler shift of the star and solar spectrum is less
than 10\,km\,s$^{-1}$ (i.e. within a line width).
This 15\,m\,s$^{-1}$ error was determined from 10 stars for which
observations were taken with and without moonlight subtraction.
The first 5 RV measurements of Kepler-18 were made with the B5 decker
and hence suffer from such errors.
Decker C2 is long enough to permit us to do accurate subtraction of
any moonlit sky spectrum.   Sky subtraction was not possible with the
spectra taken with decker B5.  
The iodine absorption cell was used as the velocity metric.
This HIRES configuration can give a velocity
precision as good as 1.0\,m\,s$^{-1}$, depending on the stellar spectral
type, the stellar rotation velocity \vsini, and on the signal-to-noise
ratio of the observation.
The exposure times for the {\koiname} spectra ranged from 2300 to 2700
seconds, and the final signal-to-noise ratios ranged from 63 to 80 per
pixel, or 127 to 162 per 4.1\,pixel resolution element.
The Doppler RV analysis algorithm \citep{BuMaWi96,JoWiAl09}
computes an uncertainty in each data point from
the variance about the mean of the individual spectral chunks into which
the spectrum is divided. Main sequence stars having Teff near 5400K have
been measured for precise
RVs for roughly 100 stars using the same HIRES instrument.  Such stars
typically show an additional noise of 2 m/s, caused by surface velocity
fields and instrumental effects.   We have modeled this additional
intrinsic short-term stellar variability by adding that 2.0\,m\,s$^{-1}$
``jitter'' in quadrature to the uncertainties of
each RV data point computed by the RV code.
The measured HIRES relative velocities are given in Table~\ref{tab:RVs}.
These velocity measurements are shown as the black points in
Figure~\ref{fig:RVs}.  The solid blue line in this figure is the model
fit from the joint MCMC solution of the RV data and the lightcurve,
presented in Section~\ref{sec:LCRV}.
The rms scatter of the RV observations around this model fit is
4.3\,m\,s$^{-1}$, whereas the rms scatter of the raw RVs is
9.6\,m\,s$^{-1}$.
The inferred planetary masses from this solution are given in
the first row of Table~\ref{tab:FinalMasses}.
Also shown as the dashed red line in Figure~\ref{fig:RVs} is a two-planet
radial velocity solution that includes only {\planetc} (\koione) and
{\planetd} (\koitwo).
The black line in Figure~\ref{fig:RVs} shows the velocities from the
full multi-body dynamical solution discussed in Section~\ref{sec:TTV}.
The masses from this dynamical solution are adopted in
Section~\ref{sec:TTV} as our best determination of the planet masses.

%
%

\begin{deluxetable}{rrr}
\tablenum{3}
\tablewidth{0pt}
\tablecaption{Keck HIRES Relative Radial Velocity Measurements of \koiname.
\label{tab:RVs}}
\tablehead{\colhead{BJD} & \colhead{RV}           & \colhead{$\sigma$} \\
                        & \colhead {m\,s$^{-1}$} & \colhead {m\,s$^{-1}$} }
\startdata
2455076.008719 &   4.67 & 5.02 \\
2455076.927037 &   4.24 & 4.99 \\
2455081.024475 &   5.43 & 8.10 \\
2455082.007257 &   1.09 & 4.70 \\
2455084.983537 &  -8.83 & 5.61 \\
2455318.066020 &   1.65 & 5.36 \\
2455322.029214 & -11.97 & 4.64 \\
2455373.003937 &  10.85 & 4.52 \\
2455403.018946 &  21.47 & 5.85 \\
2455405.909151 & -12.24 & 4.73 \\
2455406.881186 &  -0.53 & 4.39 \\
2455413.010870 & -10.44 & 5.24 \\
2455432.969613 &   1.21 & 4.45 \\
2455436.781954 &  -8.54 & 4.45 \\
\enddata
\end{deluxetable}

%
%

\subsection{Spectroscopic Analysis}\label{sec:photosphere}
We determined stellar parameters using the local thermodynamic equilibrium
(LTE) line analysis and spectral synthesis code
MOOG\footnote{available at http://www.as.utexas.edu/\~chris/moog.html}
\citep{Sn73}, together with a grid of
Kurucz\footnote{http://kurucz.harvard.edu/grids.html} ATLAS9 model atmospheres.
The method used is virtually identical to that described in \citet{BrDRCo11}.
We analyzed a spectrum of {\starname} obtained with the Keck~1 HIRES
spectrograph on 27 August 2010\,UT.  We first
measured the equivalent widths of a carefully-selected list of 53 neutral
iron lines and 13 singly-ionized iron lines in a spectrum of the Jovian
satellite Ganymede,
taken using the same instrumental setup and configuration as that used
for \starname.  MOOG force-fits abundances to match these measured equivalent
widths, using declared atomic line parameters.  By assuming excitation
equilibrium, we constrained the stellar temperature by eliminating any
trends with excitation potential; assuming ionization equilibrium, we
constrained the stellar surface gravity by forcing the derived iron
abundance using neutral lines to match that of singly-ionized lines.  The
microturbulent velocity $\xi$ was constrained by eliminating any trend with
reduced equivalent width ($\rm{W}_{\lambda}/\lambda$).
Our derived stellar parameters for the Sun (using our Ganymede spectrum)
are as follows: $\teff = 5785\pm70$\,K,
$\logg = 4.54\pm0.09$\,dex, microturbulent velocity
$\xi = 1.17\pm0.06$\,km\,s$^{-1}$, and Fe abundance
$\log(\epsilon)=7.54\pm0.05$\,dex.

We repeated the process described above for the spectrum of \starname.
We then took the difference, on a line-by-line basis, of the derived
iron abundance from each line.  Our resulting iron abundance is therefore
differential with respect to the Sun.  To estimate the rotational
velocity of \starname, we synthesized three 5{\AA}
wide spectral regions in the range 5640--5690{\AA} and adjusted
the Gaussian and rotational broadening parameters until the
best fit (by eye) was found to the observed spectrum.  The results of our
analysis yield the following stellar parameters for \starname:
$\teff = \MOOGteff$\,K, $\logg = \MOOGlogg$, $\xi = 1.09\pm0.08$\,km\,s$^{-1}$,
$\feh = \MOOGfeh$ and $\vsini \MOOGvsini$\,km\,s$^{-1}$.
The sky-projected rotational velocity of the star is very small.
For such small {\vsini} values, disentangling line-broadening
effects due to the instrument, macroturbulence and rotation
is difficult, at best, and requires higher signal-to-noise and
resolution than our spectrum offers.  In our MOOG analysis, we have
therefore chosen to quote an upper-limit for \vsini,
which we estimate by assuming {\itshape all} broadening is due
to stellar rotation.

A completely independent analysis of the same spectrum was done
using the stellar spectral synthesis package SME \citep{VaPi96,VaFi05}.
This analysis of {\starname} gives $\teff = \SMEteff$\,K, $\logg = \SMElogg$,
$\feh = \SMEfeh$ and $\vsini = \SMEvsini$\,km\,s$^{-1}$.
The agreement between the MOOG and the SME analyses of this star is
excellent.   We have arbitrarily selected the MOOG parameters as our
adopted values.

\section{Light Curve and Radial Velocity Solution}\label{sec:LCRV}

\begin{figure*}
\includegraphics[angle=-90,width=\textwidth]{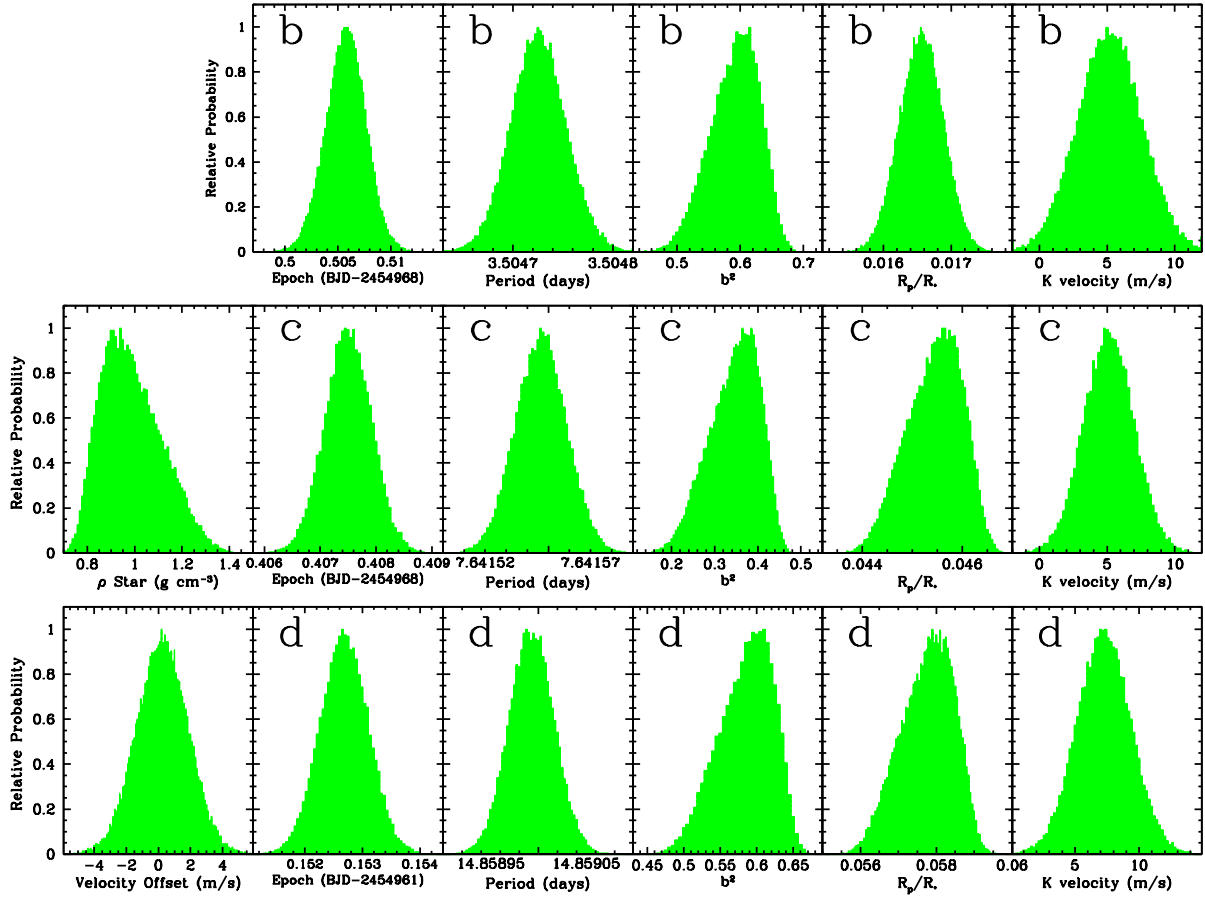}
\caption{Markov Chain Monte Carlo model probability distribution functions
for parameters of the {\starname} system.   The left column gives
the two stellar parameters of the stellar density and the relative
radial velocity zero point.  For the other five columns, the top
row gives the planet parameters of transit epoch, orbital period, square of
the impact parameter, ratio of the planetary to stellar radius, and the
orbital velocity amplitude $K$ for {\planetb}.  The second and third rows
give the same parameters for {\planetc} and {\planetd} respectively.
\label{fig:MCMC}}
\end{figure*}

We performed a joint solution of the photometric light curve and the radial
velocity data in order to derive the orbital parameters as well as the
physical characteristics of the planet candidates, in the same manner as
described in detail by \citet{BaBoBr11} for Kepler-10b.
The transit lightcurves were modeled using the analytic formulation of
\citet{MaAg02}, and the radial velocities were fit with a Keplerian orbit.
We model the mean density of the star, and for each of the three planet
candidates we model the planetary radius, the orbital period, T$_0$,
the impact parameter $b$ and the radial velocity amplitude $K$. 
The eccentricity was fixed to 0.0.
Model parameters were determined by minimizing the $\chi^2$ statistic using
a Levenberg-Marquardt technique.   To determine the best-fit model
parameter distributions, we used a Markov Chain Monte Carlo (MCMC) method
that has been optimized for highly correlated parameters \citep{Gr11a}.
Stellar parameters were determined in a separate MCMC analysis using the
Yonsei-Yale isochrones \citep[][et seq.]{YiDeKi01} with \teff, {\feh} and
fitted value of the mean-stellar density as constraints.
Figure~\ref{fig:MCMC} shows the parameter distribution functions
resulting from the MCMC analysis. 
Table~\ref{tab:SystemParameters} lists the adopted system parameters.
In most cases, these parameters came from the MCMC analysis.
As discussed in Section~\ref{sec:TTV}, significant transit timing
variations were detected in {\planetc} and {\planetd}.  These
transit timing variations would have been absorbed into the MCMC
periods and epochs and their uncertainties.  Therefore we have adopted the
periods and epochs from the TTV analysis, and these are reported in
Table~\ref{tab:SystemParameters}. The uncertainty in the
epoch is the median absolute deviation of the transit times from this
ephemeris and the uncertainty in the period is this quantity divided
by the number of orbits between the first and last observed transits.
Similarly, the dynamical system model presented in Section~\ref{sec:TTV}
derives our best and most reliable masses for the three planets.
These masses from the dynamical model are reported in
Table~\ref{tab:SystemParameters}.  The masses from the MCMC model
are listed in Table~\ref{tab:FinalMasses}.
The TTV dynamical model places tight upper limits on the orbital
eccentricity of planets "c" and "d", as shown in
Table~\ref{tab:TTVsolution}.  These eccentricities are small enough that
they would not significantly change the parameters in
Table~\ref{tab:SystemParameters}.  For example, the zero eccentricity
solution gives a stellar density of $1.01\pm0.12$\,g/cm$^3$.
Inclusion of the eccentricity from the TTV solution would change this
by $0.03$\,g/cm$^3$, or $0.25\sigma$.
For each parameter adopted from the MCMC model, we adopt the median
value of the distribution, and the error bar represents the 68.3\%
confidence level -- roughly equivalent to a $1\sigma$ confidence level.

It is interesting and informative to compare the initial values for the
transiting planet parameters presented in the tabulation of candidates by
\citet{BoKoBa11b} with those we have finally settled on in
Table~\ref{tab:SystemParameters}.  The parameters of the parent star given
by \citet{BoKoBa11b} {\koiname} are simply taken from the KIC.  The overall
accuracy of the KIC photometric stellar parameters is actually quite good,
as was shown by \citet{BrLaEv11}.  The other important planetary system
parameters reported by \citet{BoKoBa11b} are the orbital periods and
transit depths of each of the three KOIs in the system.   Our adopted
orbital periods agree with the preliminary \citet{BoKoBa11b} values to
within 1\,$\sigma$ for {\koione} and {\koithree}, and within 4\,sigma for
{\koitwo}.   We note that \citet{BoKoBa11b} assumed no transit time
variations, and fit the best mean period to the data.   
The transit depths agree to within 2\,$\sigma$ for {\koitwo} and
{\koithree}, but differ by 7.5\,$sigma$ for {\koitwo}.  This slight
disagreement for {\koitwo} probably results from \citet{BoKoBa11b} fitting an
impact parameter of 0.03, whereas our best value is $0.593 \pm 0.050$.

The radial velocity variations of the parent star {\starname} resulting from
this three planet MCMC model is shown as the solid blue line in
Figure~\ref{fig:RVs}.
This solution is very similar to the three-body radial velocity solution one
would get from the radial velocity data alone, holding the period and $T_0$
of each planet fixed at the values from the photometric lightcurve solution
and assuming zero eccentricity.
A two-body solution to the radial velocity data considering only {\koione}
and {\koitwo} is shown as the dashed red line in Figure~\ref{fig:RVs}.
The full three-planet solution gives a significantly better fit to the
observed radial velocities than does the two-body solution.  The rms of
the full MCMC three-body fit is 4.3\,m\,s$^{-1}$, while the rms of
the two-body Keplerian solution is 5.5\,m\,s$^{-1}$.  We computed an $f$-test
based on the residuals of a 2 planet and 3-planet RV fit. 
The periods where fixed based on transit data.  We find $f=4.97$, which means
there is a 0.68\% probability that variance of the residuals are similar.
Thus this reduction
in the rms supports the interpretation of {\koithree} as a third planetary
companion to \koiname.   However, the radial velocity observations are not
sufficiently dense to consider this to be a confirmation of this KOI as a
planet.

%
%

\begin{deluxetable}{lc}
\tablenum{4}
\tighten
\tablewidth{0pt}
\tablecaption{Adopted System Parameters
\label{tab:SystemParameters}}
\tablehead{\colhead{ Parameter}  & \colhead{Value}}
\startdata
\mstar\ (\msun)\tablenotemark{a}	&   0.972 $\pm$ 0.042\\
\rstar\ (\rsun)\tablenotemark{a}	&   1.108 $\pm$ 0.051\\
\loglstar\ (\lsun)\tablenotemark{a}	&  -0.031  $\pm$ 0.035\\
Age (Gyr)\tablenotemark{a}		&  10.0   $\pm$ 2.3 \\
\logg$_\star$\tablenotemark{b}		&   \MOOGlogg \\
\rhostar\ (g/cm$^3$)\tablenotemark{c}	&   1.01  $\pm$ 0.12\\
\hline
\multicolumn{2}{c}{{\planetb} = \koithree} \\
$T_0$\tablenotemark{c}			& \epochb		\\
$P$ (days)\tablenotemark{c}		& \periodb		\\
Transit depth (ppm)\tablenotemark{c}	& $ 254.0 \pm 7.8 $	\\
$b$ (Impact Parameter)\tablenotemark{c}	& \impactb		\\
$R_p$/{\rstar}\tablenotemark{c}		& \scaledPlanetRadiusb	\\
$M_p$ (\mearth)\tablenotemark{d}	& \TTVmplanetb		\\
$R_p$ (\rearth)\tablenotemark{c}	& \rplanetb		\\
$i$ (deg)\tablenotemark{e}		& $84.92   \pm 0.26 $	\\
$a/{\rstar}$\tablenotemark{e}		& $ 8.58   \pm 0.37 $	\\
$\bar{\rho}_p$ (g/cm$^3$)\tablenotemark{3,4} & \TTVrhoplanetb	\\
$a$ (AU)\tablenotemark{e}		& $ 0.0447 \pm 0.0006 $	\\
$K$ (m/s)\tablenotemark{c}		& $ 5.2    \pm 2.4    $	\\
$T_{14}$ (h)\tablenotemark{e}		& $ 2.076  \pm 0.036  $	\\
$T_{12}$ (h)\tablenotemark{e}		& $ 0.0818 \pm 0.0082 $	\\
\hline
\multicolumn{2}{c}{{\planetc} = {\koione}} \\
$T_0$\tablenotemark{d}       		& \TTVepochc		\\
$P$ (days)\tablenotemark{d}		& \TTVperiodc		\\
Transit depth (ppm)\tablenotemark{c}	& $ 2286.6 \pm 8.6 $	\\
$b$ (Impact Parameter)\tablenotemark{c}	& \impactc		\\
$R_p$/{\rstar}\tablenotemark{c}		& \scaledPlanetRadiusc	\\
$M_p$ (\mearth)\tablenotemark{d}	& \TTVmplanetc		\\
$R_p$ (\rearth)\tablenotemark{c}	& \rplanetc		\\
$i$ (deg)\tablenotemark{e}		& $87.68   \pm 0.22  $	\\
$a/{\rstar}$\tablenotemark{e}		& $14.43   \pm 0.61  $	\\
$\bar{\rho}_p$ (g/cm$^3$)\tablenotemark{3,4}   & \TTVrhoplanetc	\\
$a$ (AU)\tablenotemark{e}		& $ 0.0752 \pm 0.0011 $	\\
$K$ (m/s)\tablenotemark{c}		& $ 5.1    \pm 1.9    $	\\
$T_{14}$ (h)\tablenotemark{e}		& $ 3.488  \pm 0.020  $	\\
$T_{12}$ (h)\tablenotemark{e}		& $ 0.229  \pm 0.022  $	\\
\hline
\multicolumn{2}{c}{{\planetd} = {\koitwo}} \\
$T_0$\tablenotemark{d}			& \TTVepochd		\\
$P$ (days)\tablenotemark{d}		& \TTVperiodd		\\
Transit depth (ppm)\tablenotemark{c}	& $ 3265. \pm 12. $	\\
$b$ (Impact Parameter)\tablenotemark{c}	& \impactd		\\
$R_p$/{\rstar}\tablenotemark{c}		& \scaledPlanetRadiusd	\\
$M_p$ (\mearth)\tablenotemark{d}	& \TTVmplanetd		\\
$R_p$ (\rearth)\tablenotemark{c}	& \rplanetd		\\
$i$ (deg)\tablenotemark{e}		& $ 88.07  \pm 0.10 $	\\
$a/{\rstar}$\tablenotemark{e}		& $ 22.48  \pm 0.96 $	\\
$\bar{\rho}_p$ (g/cm$^3$)\tablenotemark{3,4}  & \TTVrhoplanetd	\\
$a$ (AU)\tablenotemark{e}		& $ 0.1172 \pm 0.0017$	\\
$K$ (m/s)\tablenotemark{c}		& $ 7.3    \pm 2.1   $	\\
$T_{14}$ (h)\tablenotemark{e}		& $ 3.679  \pm 0.036 $	\\
$T_{12}$ (h)\tablenotemark{e}		& $ 0.459  \pm 0.045 $	\\
\enddata
\tablenotetext{a}{based on isochrone fits using {\rhostar}, from MCMC model
and {\teff} and {\feh} from spectroscopy.}
\tablenotetext{b}{from MOOG analysis}
\tablenotetext{c}{from MCMC model}
\tablenotetext{d}{from TTV model}
\tablenotetext{e}{derived from other parameters}
\end{deluxetable}

%
%

\section{\wspitzer\ Observations of {\koione} and {\koitwo}}
\label{sec:spitzer}

Observation of the transits of the planets around {\starname} at multiple,
widely-spaced wavelengths is a valuable tool to confirm the planetary
nature of the events.  The depth of a planetary transit should be nearly
independent of wavelength, aside from minor effects due to the possible
finite brightness of the planet as a function of wavelength.
In order to test for wavelength independence of the transit depths,
{\koione} and {\koitwo} were observed during two transits with \wspitzer/IRAC
\citep{WeRoLo04,FaHaAl04} at 4.5~\micron\ (program ID 60028).
The observations occurred on UT 2010 July 19 and UT 2010 August 13.
Both visits lasted approximately 9\,h.
The data were gathered in full-frame mode ($256\times256$ pixels)
with an exposure time of 12\,s per image which yielded 2418 and 2575
images per respective visit.

The method we used to produce photometric time series from the images is
described by \citet{DeEtHe09}.
It consists of finding the centroid position of the stellar point spread
function (PSF) and performing aperture photometry using a circular aperture

\begin{figure*}
\includegraphics[width=0.48\textwidth]{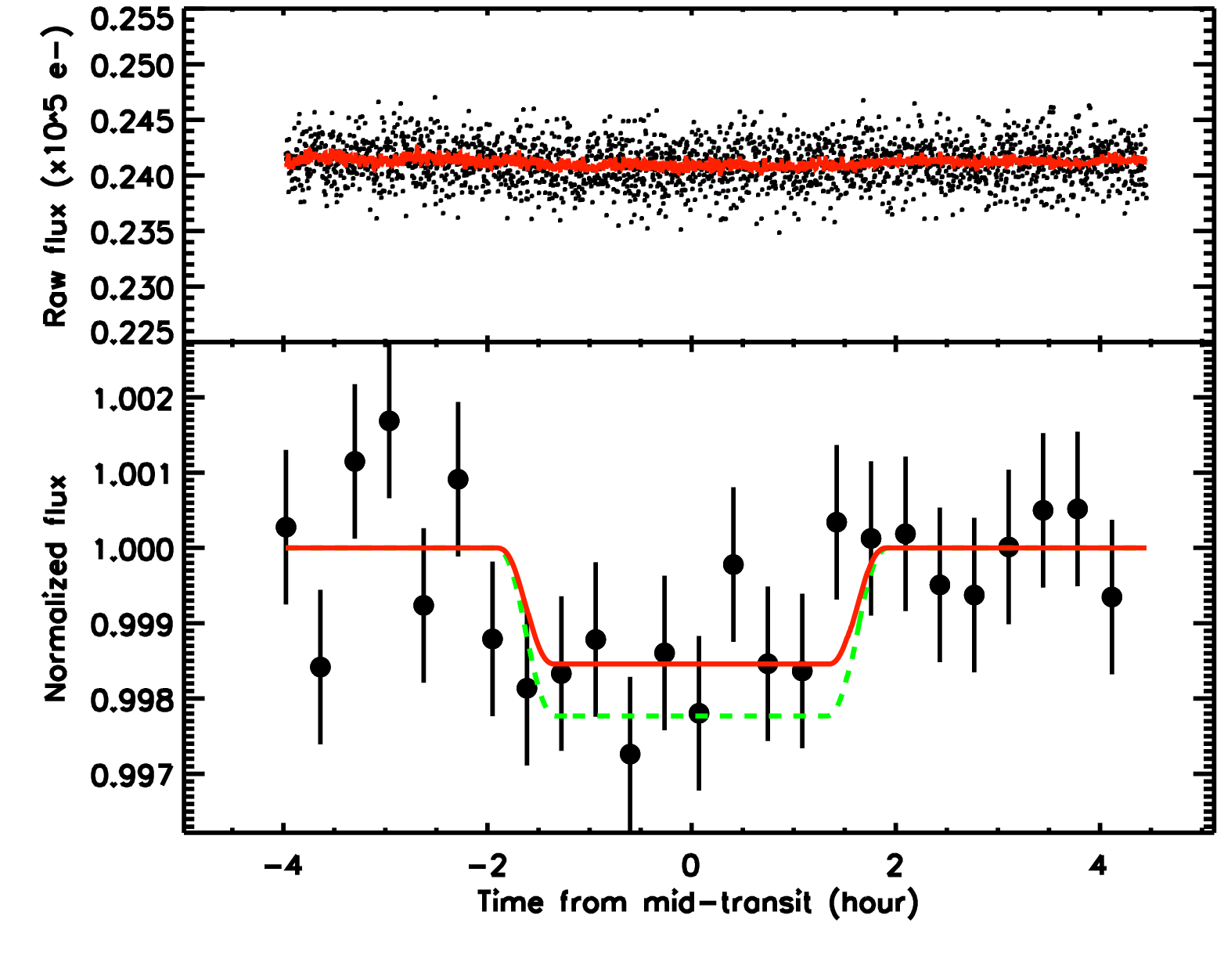}
\includegraphics[width=0.48\textwidth]{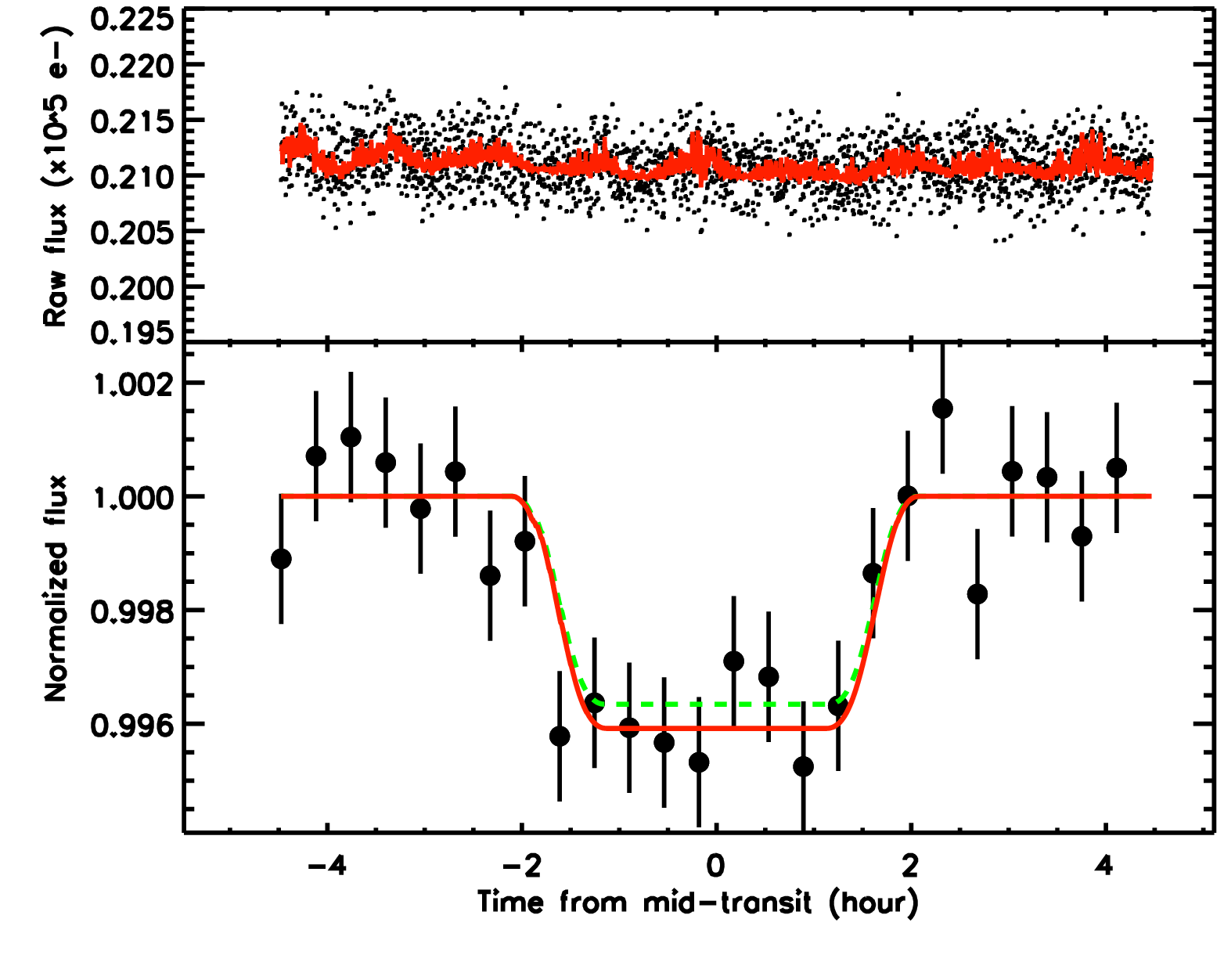}
\caption{{\spitzer} transit light-curves of {\koione} (top) and  {\koitwo}
(bottom) observed in the IRAC band-pass at 4.5~\micron.
Top panels~: raw (unbinned) transit light-curves.
The red solid lines correspond to the best fit models which include
the time and position instrumental decorrelations as well as the model
for the planetary transit (see details in Sect.~\ref{sec:spitzer}).
Lower panels~: transit light-curve  corrected, normalized and binned
by 36 minutes.  The best-fit {\spitzer} transit curves are plotted in
red and the transit shapes expected from the {\kepler} observations are
overplotted in dashed green lines.}
\label{fig:spitzerlightcurves}
\end{figure*}

The images used are the Basic Calibrated Data (BCD) delivered by the
\emph{Spitzer} archive.
These files are corrected for dark current, flat-fielding, detector
non-linearity and converted into flux units.
We convert the pixel intensities to electrons using the information given
in the detector gain and exposure time provided in the FITS headers.
This facilitates the evaluation of the photometric errors.
We extract the UTC-based Julian date for each image from the FITS header
(keyword DATE\_OBS) and correct to mid-exposure.
We convert to UTC-based BJD following the procedure developed by
\citet{EaSiGa10}.
We use the JPL Horizons ephemeris to estimate the position of \spitzer\
Space Telescope during the observations.
We correct for transient pixels in each individual image using a 20-point
sliding median filter of the pixel intensity versus time.
For this step, we compare each pixel's intensity to the median of the 10
preceding and 10 following exposures at the same pixel position and we
replace outliers greater than $4 \sigma$ with their median value.
The fraction of pixels we correct is lower than 0.16\% for both transits.
The centroid position of the stellar PSF is determined using DAOPHOT-type
Photometry Procedures, \texttt{GCNTRD}, from the IDL Astronomy Library
\footnote{{\tt http://idlastro.gsfc.nasa.gov/homepage.html}}. We use the
\texttt{APER} routine to perform aperture photometry with a circular
aperture of variable radius, using radii of 1.5 to 8 pixels, in 0.5 pixel steps.
The propagated uncertainties are derived as a function of the aperture
radius; we adopt the one which provides the smallest errors.
We find that the transit depths and errors vary only weakly with the
aperture radius for all the light-curves analyzed in this project.
The optimal apertures is found to be at $3.0$~pixels.
We estimate the background by fitting a Gaussian to the central region of
the histogram of counts from the full array.
The center of the Gaussian fit is adopted as the residual background intensity.
As already seen in previous \wspitzer\ observations
\citep{DeKnAg11,BeKnBu11}, we find that the background varies by 20\% between
three distinct levels from image to image, and displays a ramp-like
behavior as function of time.
The contribution of the background to the total flux from the star is low
for both observations, from 0.15\% to 0.9\% depending on the images.
Therefore, photometric errors are not dominated by fluctuations in the
background.
We used a sliding median filter to select and trim outliers in flux and
position greater than $5~\sigma$, which corresponds to $0.9\%$ and  $2.0\%$
of the data, for the first and second visit respectively.
We also discarded the first half-hour of observations, which are affected
by a significant telescope jitter before stabilization.
The final number of photometric measurements used is $2246$ and $2369$.
The raw time series are presented in the top panels of
Figure~\ref{fig:spitzerlightcurves}.
We find that the typical signal-to-noise ratio ($S/N$) $140$ per image
which corresponds to 80\% of the theoretical signal-to-noise.
Therefore, the noise is dominated by Poisson photon noise.

%
%

\subsection{Analysis of the \wspitzer\ light curves}

We used a transit light curve model multiplied by instrumental
decorrelation functions to measure the transit parameters and their
uncertainties from the \spitzer\ data as described in \cite{DeSiVM11}.
We compute the transit light curves with the IDL transit routine
\texttt{OCCULTSMALL} from \cite{MaAg02}.
In the present case, this function depends on one
parameter: the planet-to-star radius ratio $R_p / R_\star$.
The orbital semi-major axis to stellar radius ratio (system scale)
$a / R_\star$, the impact parameter $b$, and the time of mid transit
T$_0$ are fixed to the values derived from the {\kepler} lightcurves.
The limb-darkening coefficients are set to zero since these \spitzer\
lightcurves do not have enough photometric precision to detect the curvature
in the transit light curve that would be produced by limb darkening.

The \spitzer/IRAC photometry is known to be systematically affected by the
so-called \textit{pixel-phase effect} (see e.g., \citealt{ChAlMe05,KnChAl08}).
This effect is seen as oscillations in the measured fluxes with a period of
approximately 70~min (period of the telescope pointing jitter) and an
amplitude of approximately $2\%$ peak-to-peak.
We decorrelated our signal in each channel using a linear function of time
for the baseline (two parameters) and a quadratic function of the PSF
position (four parameters) to correct the data for each channel.
We checked that adding parameters to the correction function of the PSF
position \citep[as in][]{DeEtHe09} does not improve the fit significantly.
We performed a simultaneous Levenberg-Marquardt least-squares fit
\citep{Ma09} to the data to determine the transit and instrumental
model parameters (7 in total).
The errors on each photometric point were assumed to be identical, and were
set to the $rms$ of the residuals of the initial best-fit obtained.
To obtain an estimate of the correlated and systematic errors
\citep{PoZuQu06} in our measurements, we use the residual permutation
bootstrap, or ``Prayer Bead'', method as described in \citet{DeEtHe09}. In
this method, the residuals of the initial fit are shifted systematically
and sequentially by one frame, and then added to the transit light curve
model before fitting again.
We allow asymmetric error bars spanning $34\%$ of the points above and
below the median of the distributions to derive the $1~\sigma$
uncertainties for each parameters as described in \citet{DeChFo11}.

We measured the transit depths at 4.5~\micron\ of
$1521^{+277}_{-245}$ ppm for {\planetc} and
$4083^{+298}_{-285}$ for {\planetd}.
The values we measure for the transit depths in the \spitzer\ bandpass are
in agreement at the 2$\sigma$ level compared to the {\kepler} bandpass.
This indicates that the transit depths of {\planetc} and {\planetd} are only
weakly dependent on wavelength, if at all. 
This is in agreement  with expectations for a dark planetary object, and
indicates that there is no significant contamination from any nearby
unresolved star of significantly different color.
As discussed in Section~\ref{sec:blender},
this is important evidence that the transit signals arise from planets
and not from eclipsing stars blended with additional light.

These Spitzer observations provide a useful constraint on the kinds of false
positives (blends) that may be mimicking the transit signal, such as
eclipsing binaries blended with the target.
If {\starname} were blended with an unresolved eclipsing binary of later
spectral type that manages to reproduce the transit depth
in the Kepler passband, the predicted depth at 4.5$\mu$m would be expected to
be larger because of the higher flux of the contaminant at longer
wavelengths compared to {\starname}.
Since the transit depth we measure in the near infrared is about the same
as in the optical, this argues against blends composed of stars of much
later spectral type.
Based on model isochrones, the properties of the target star, and the
transit depths measured with Spitzer at the $3\sigma$ level,
we determine a lower limit to the blend masses of 0.86 and $0.79\msun$,
for {\planetc} and {\planetd} respectively.

\section{Transit Timing Variations Analysis}\label{sec:TTV}

\subsection{Transit Times and Errors}

{\starname} was observed at the 29.4244~minute Long Cadence (LC) rate
for the first three observing periods Q0-Q2.  
After the transits of {\koione} and
{\koitwo} were detected, {\starname} was observed at the 58.85~second
Short Cadence
in order to facilitate the measurement of possible transit timing
variations.   These SC data were used for Q3 through Q7.
Transit times and their errors for each member of the {\starname}
system were determined through an iterative procedure;
a single step of this procedure is described as follows.

First, the detrended photometric data within four transit durations
at each epoch were shifted by the current best-fit mid-transit times,
and the lightcurves were folded on these transit times to form
a transit template.  This template was then fit with a transit
light curve model \citep{MaAg02}.
Next, at each individual epoch, this light curve template was shifted
in time and compared to the data by computing the standard $\chi^2$
statistic.  This statistic was computed for a dense, uniform sample of
mid-transit times centered on the current best-fit time and spanning
approximately five current best-fit timing errors.  The time $t_0$,
corresponding to the minimum $\chi^2$, was recorded as the new
best-fit mid-transit time.  The $\chi^2$ data were then fit with a
quadratic function of time, $\chi^2(t) = C (t-t_0)^2+\chi_0^2$.  By
choosing this functional form, we are assuming that the posterior
likelihood is well-described by a Gaussian function of the mid-transit
time. The fidelity of the quadratic approximation was verified
visually at each epoch. The timing error, $\sigma_t$, was found by
solving for the time, $t = t_0+\sigma_t$, at which the quadratic model
indicated a $\Delta \chi^2$ = 1.  This corresponds to $\sigma_t =
C^{-1/2}$.

This iterative procedure of template generation followed by timing
estimate converged to the final values in generally two to three
steps.  The measured transit times for each observed transit of
{\planetc} are given in Table~\ref{tab:TT01}, and the measured transit
times for {\planetd} are given in Table~\ref{tab:TT02}.

\begin{deluxetable}{rrrr}
\tablenum{5}
\tablewidth{0in}
\tablecaption{Transit Times for \planetc\ = K00137.01 \label{tab:TT01}}
\tablehead{
\colhead{cycle\tablenotemark{a}}  &
\colhead{BJD-2455000.0} &
\colhead{O-C} &
\colhead{$\sigma$}}
\startdata
-27.0& -39.23132& $+$0.00343& 0.00141\\
-26.0& -31.59238& $+$0.00077& 0.00128\\
-25.0& -23.94944& $+$0.00212& 0.00134\\
-24.0& -16.30829& $+$0.00168& 0.00125\\
-23.0&  -8.66703& $+$0.00134& 0.00202\\
-21.0&   6.61139& $-$0.00342& 0.00161\\
-20.0&  14.25400& $-$0.00241& 0.00155\\
-19.0&  21.89543& $-$0.00257& 0.00149\\
-18.0&  29.53633& $-$0.00327& 0.00135\\
-17.0&  37.17764& $-$0.00355& 0.00131\\
-16.0&  44.82280& $+$0.00002& 0.00146\\
-15.0&  52.46001& $-$0.00436& 0.00107\\
-14.0&  60.10428& $-$0.00169& 0.00115\\
-13.0&  67.74533& $-$0.00223& 0.00108\\
-12.0&  75.39019& $+$0.00103& 0.00145\\
-11.0&  83.02837& $-$0.00238& 0.00144\\
-10.0&  90.67043& $-$0.00191& 0.00140\\
 -9.0&  98.31334& $-$0.00059& 0.00102\\
 -8.0& 105.95509& $-$0.00044& 0.00105\\
 -7.0& 113.59742& $+$0.00030& 0.00148\\
 -6.0& 121.23989& $+$0.00117& 0.00098\\
 -5.0& 128.88188& $+$0.00157& 0.00108\\
 -4.0& 136.52252& $+$0.00062& 0.00102\\
 -3.0& 144.16670& $+$0.00320& 0.00088\\
 -2.0& 151.80782& $+$0.00273& 0.00137\\
 -1.0& 159.44986& $+$0.00318& 0.00113\\
  0.0& 167.09286& $+$0.00459& 0.00096\\
  1.0& 174.73457& $+$0.00470& 0.00097\\
  2.0& 182.37715& $+$0.00568& 0.00131\\
  3.0& 190.01671& $+$0.00365& 0.00106\\
  4.0& 197.65742& $+$0.00277& 0.00109\\
  5.0& 205.30025& $+$0.00400& 0.00117\\
  6.0& 212.94064& $+$0.00281& 0.00101\\
  7.0& 220.57949& $+$0.00006& 0.00111\\
  8.0& 228.22281& $+$0.00178& 0.00101\\
  9.0& 235.86365& $+$0.00103& 0.00106\\
 10.0& 243.50537& $+$0.00116& 0.00107\\
 11.0& 251.14597& $+$0.00017& 0.00097\\
 12.0& 258.78515& $-$0.00224& 0.00092\\
 13.0& 266.42783& $-$0.00116& 0.00115\\
 14.0& 274.06799& $-$0.00259& 0.00118\\
 15.0& 281.70882& $-$0.00335& 0.00110\\
 16.0& 289.35100& $-$0.00277& 0.00079\\
 17.0& 296.99249& $-$0.00288& 0.00088\\
 18.0& 304.63304& $-$0.00392& 0.00083\\
 19.0& 312.27505& $-$0.00350& 0.00093\\
 20.0& 319.91502& $-$0.00512& 0.00090\\
 21.0& 327.56008& $-$0.00165& 0.00089\\
 22.0& 335.19905& $-$0.00427& 0.00093\\
 23.0& 342.84138& $-$0.00355& 0.00084\\
 24.0& 350.48436& $-$0.00215& 0.00103\\
 25.0& 358.12585& $-$0.00226& 0.00097\\
 26.0& 365.76604& $-$0.00367& 0.00086\\
 27.0& 373.41078& $-$0.00052& 0.00095\\
 28.0& 381.05287& $-$0.00001& 0.00101\\
 29.0& 388.69546& $+$0.00098& 0.00089\\
 30.0& 396.33836& $+$0.00228& 0.00090\\
 31.0& 403.97897& $+$0.00130& 0.00096\\
 32.0& 411.62239& $+$0.00313& 0.00104\\
 33.0& 419.26374& $+$0.00288& 0.00090\\
 34.0& 426.90502& $+$0.00256& 0.00100\\
 35.0& 434.54873& $+$0.00468& 0.00112\\
 36.0& 442.18944& $+$0.00381& 0.00091\\
 37.0& 449.82907& $+$0.00184& 0.00106\\
 38.0& 457.46935& $+$0.00052& 0.00097\\
 39.0& 465.11288& $+$0.00246& 0.00092\\
 40.0& 472.75469& $+$0.00268& 0.00103\\
 41.0& 480.39522& $+$0.00161& 0.00117\\
 42.0& 488.03539& $+$0.00020& 0.00101\\
 43.0& 495.67681& $+$0.00002& 0.00100\\
 44.0& 503.31876& $+$0.00037& 0.00092\\
 45.0& 510.95872& $-$0.00125& 0.00091\\
 46.0& 518.60144& $-$0.00014& 0.00088\\
 47.0& 526.24067& $-$0.00249& 0.00114\\
 48.0& 533.88235& $-$0.00241& 0.00112\\
\enddata
\tablenotetext{a}{$P =7.64159$ days, $T_0 = 2455167.08828$}
\end{deluxetable}

\begin{deluxetable}{rrrr}
\tablenum{6}
\tablewidth{0in}
\tablecaption{Transit Times for \planetd\ = K00137.02 \label{tab:TT02}}
\tablehead{
\colhead{cycle\tablenotemark{a}}  &
\colhead{BJD-2455000.0} &
\colhead{O-C} &
\colhead{$\sigma$}}
\startdata
-14.0& -38.84989& $-$0.00318& 0.00101\\
-13.0& -23.98927& $-$0.00144& 0.00090\\
-12.0&  -9.12888& $+$0.00008& 0.00081\\
-11.0&   5.73040& $+$0.00048& 0.00086\\
-10.0&  20.59242& $+$0.00362& 0.00115\\
 -9.0&  35.44894& $+$0.00126& 0.00074\\
 -8.0&  50.30774& $+$0.00119& 0.00091\\
 -7.0&  65.16644& $+$0.00101& 0.00078\\
 -6.0&  80.02557& $+$0.00126& 0.00121\\
 -5.0&  94.88445& $+$0.00127& 0.00068\\
 -4.0& 109.74201& $-$0.00005& 0.00065\\
 -3.0& 124.59986& $-$0.00107& 0.00080\\
 -2.0& 139.45743& $-$0.00238& 0.00065\\
 -1.0& 154.31562& $-$0.00307& 0.00067\\
  0.0& 169.17495& $-$0.00261& 0.00070\\
  2.0& 198.89361& $-$0.00170& 0.00081\\
  3.0& 213.75079& $-$0.00341& 0.00085\\
  4.0& 228.61267& $-$0.00040& 0.00079\\
  5.0& 243.47039& $-$0.00155& 0.00081\\
  6.0& 258.33093& $+$0.00011& 0.00068\\
  7.0& 273.19074& $+$0.00104& 0.00071\\
  8.0& 288.05139& $+$0.00282& 0.00069\\
  9.0& 302.91049& $+$0.00304& 0.00061\\
 10.0& 317.77010& $+$0.00377& 0.00065\\
 11.0& 332.62850& $+$0.00330& 0.00061\\
 12.0& 347.48637& $+$0.00229& 0.00069\\
 13.0& 362.34423& $+$0.00127& 0.00072\\
 14.0& 377.20286& $+$0.00103& 0.00070\\
 15.0& 392.06096& $+$0.00025& 0.00063\\
 16.0& 406.91891& $-$0.00068& 0.00061\\
 17.0& 421.77755& $-$0.00091& 0.00061\\
 18.0& 436.63489& $-$0.00245& 0.00063\\
 19.0& 451.49447& $-$0.00175& 0.00067\\
 20.0& 466.35351& $-$0.00158& 0.00063\\
 21.0& 481.21293& $-$0.00104& 0.00066\\
 22.0& 496.07232& $-$0.00052& 0.00067\\
 23.0& 510.93180& $+$0.00008& 0.00061\\
\enddata
\tablenotetext{a}{$P = 14.85888$~days, $T_0 = 2455169.17756$}
\end{deluxetable}

A drift in the orbital inclination was measured for each planet by
augmenting the nominal transit model with an epoch-dependent linear
inclination (viz., $i(E) = i(0)+\left(\Delta i/\Delta E\right)
\times E$), and then fitting for the linear
coefficient.  In this fit, transit times were fixed to their best-fit
values while the remaining transit parameters were allowed to vary.
The best-fitting solution was found by minimizing the standard
$\chi^2$ metric. The uncertainty was estimated by fitting a
multivariate Gaussian to a sampling of the posterior parameter
distribution.  The drifts $\left(\Delta i/\Delta E\right)$ were found
to be [$10 \pm 9$, $4 \pm 22$, $-3 \pm 10$] $\times 10^{-4}$
degrees per epoch for planets b, c, and d respectively.  Thus we
detect no secular inclination drift.

%
%
%

\subsection{Transit Time Variation Analysis}

To model the transit times and radial velocities, we applied the numerical
routines that were previously used for Kepler-9 and -11
\citep{Fa10,HoFaRa10,LiFaFo11}.  That is, a Levenberg-Marquardt
$\chi^2$-minimization routine drove 3-planet dynamical integrations, which
calculated the transit times resulting from given orbital parameters.
As in \citet{LiFaFo11}, we chose the parameters
($m_p$, $P$, $T_0$, $e \cos \omega$, $e \sin \omega$)
for each planet: mass, orbital period, transit
phase, and eccentricity vector components.
These parameters are osculating Jacobian orbital elements at
the epoch 2455168.0 [BJD].  As in \citet{HoFaRa10}, we also
used the integrations to calculate radial velocities of the
star at each of the observed times.  We assumed $M_\star = 0.95 M_\odot$,
assumed the orbits are edge-on and coplanar, and neglected light travel
time effects in these numerical calculations.

The main data constraining the orbital model are the transit times of
Tables~\ref{tab:TT01} (75 data points) and~\ref{tab:TT02} (37 data points).
In some of the calculations reported below, we allow {\koithree} to interact
dynamically, but we include only its first and last observed transits,
at $t=2454955.99237\pm0.00823$ and $2455530.77178\pm0.00174$ (2 data points),
as a way of keeping its period and phase fixed at observed values.
We have measured all of its transit times, but we do not report or
analyze them here: their individual signal to noise is low, they are
not apparently constant, and we have not found a
consistent dynamical solution to date.
Perhaps future work will show a fourth (non-transiting) planet is
required to fit the transit times of {\planetb}. In the
meantime we proceed with fitting the three transiting planets with
a focus on the transit timing constraints for {\planetc} and {\planetd}.

\begin{figure*}
\plotone{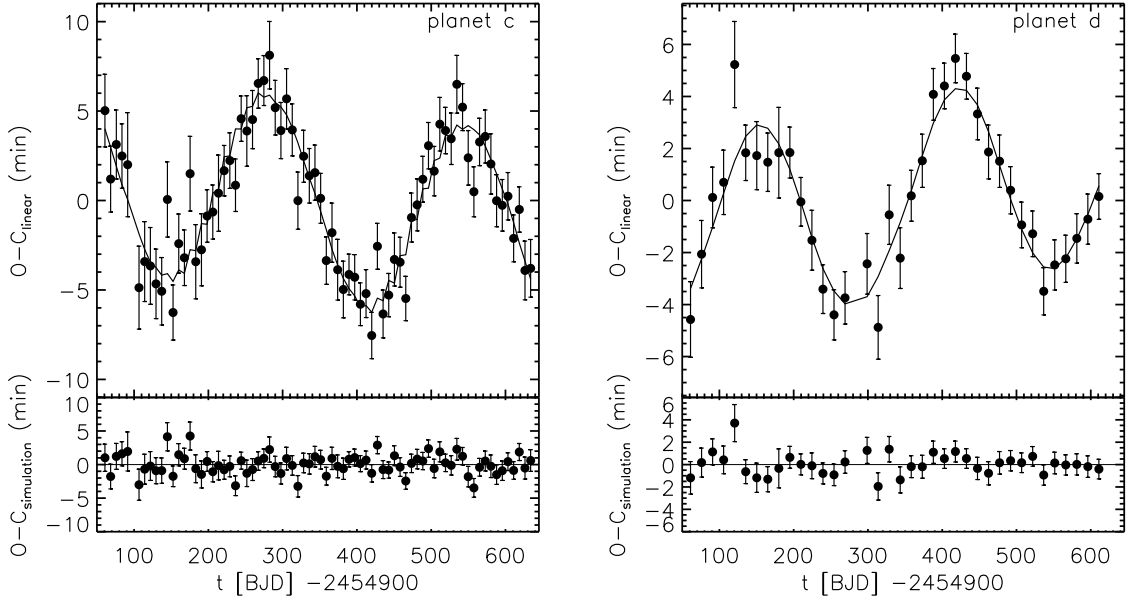}
\caption{
The observed minus calculated (based on a linear ephemeris) values of
transit times, for {\planetc} (left) and {\planetd} (right). 
The solid line shows the transit times calculated using a
dynamical model.
The lower panels show the residuals of the measurements from the model.
\label{fig:omc}}
\end{figure*}

\begin{figure}
\epsscale{1.1}
\plotone{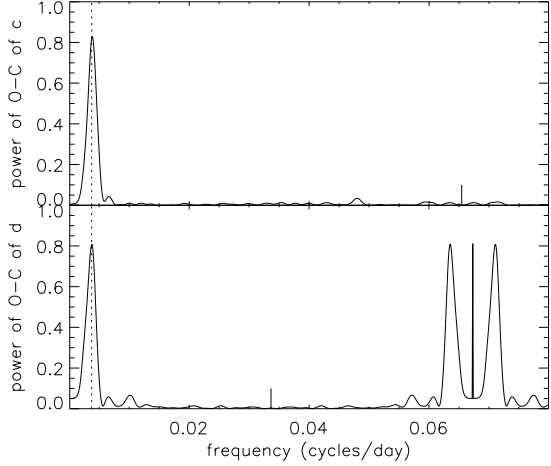}
\caption{
Periodograms \citep[fractional $\chi^2$ reduction as a function of
frequency;][]{ZeKu09} of the O-C values shown in
Figure~\ref{fig:omc}.  The dashed line shows the expected
timescale from equation~\ref{eqn:pttv}, showing excellent agreement. 
The large tick marks
show the Nyquist frequency $1/(2 P_{\rm orbital})$, beyond which the
periodogram holds no additional information.
\label{fig:periodogram}}
\end{figure}

By fitting only the transit times, allowing $P$ and T$_0$ for each of the
three planets to vary (6 free parameters), and setting dynamical
interactions to zero, we find $\chi^2=750.6$ for 114 transit time
measurements.  This is clearly an unacceptable fit, and the obvious
timing patterns call for a dynamical model. 
The observed (``O'') residuals to this calculated
(``C'') ephemeris are called the O-C values, and are plotted in
Figure~\ref{fig:omc} along with the preferred dynamical model described
below.  As in the case of Kepler-11b/c, we clearly see that the source of the
variations of transit times for {\planetc} and {\planetd} is their near 2:1
resonance.  The expected variation occurs on a timescale:
\begin{equation}
P_{TTV} = 1/(2/P_d-1/P_c) = 268{\rm\,days}, \label{eqn:pttv}
\end{equation}
the time it takes the line of conjunctions to sweep around inertial space,
through both the line of sight and the apsidal lines of these planets (on
the approximation that precession can be ignored on this timescale);
see \citet{AgStSa05}.
In figure \ref{fig:periodogram}, we plot the periodogram
of the O-C values for {\planetc} and {\planetd}.
The peaks occur at $1/(260.4 \pm 3.3\,\mathrm{days})$ and
$1/( 265.1 \pm 4.7\,\mathrm{days})$ for {\planetc} and {\planetd}
respectively, very close to the simple expectation given above, which
uniquely identifies the dynamical mechanism for transit timing variations.

Before moving on to a full solution, we tried fitting the radial velocities
(14 data points, Table~\ref{tab:RVs}) with the above-determined periods and
phases, with circular orbits, allowing only the planetary masses to vary.
The solution, listed in the second row of Table~\ref{tab:FinalMasses},
was [$m_b, m_c, m_d$] = [$12 \pm 5$, $15 \pm 5$, $28 \pm 7$]\mearth;
the $\chi^2=9.7$ for 10 degrees of freedom (14 radial velocity data points,
minus 3 K-amplitudes, minus a constant radial-velocity offset).   These masses
are well within the $1 \sigma$ error bars of the MCMC solution to the
lightcurve and RVs presented in Section~\ref{sec:LCRV}.  Therefore this
model is sufficient to explain the radial velocities, and each planet is
detected, but only marginally so for planet b.  In particular, if we hold
the mass of b at zero, the masses of the others become [$m_c$, $m_d$]=[$18
\pm 5$, $24 \pm 7$]~{\mearth} and the $\chi^2 = 15.6$ for 10 degrees
of freedom.  These values are within about 0.5$\sigma$ of the masses
determined from the joint MCMC and RV solution presented earlier in
Section~\ref{sec:LCRV}.

We may also attempt to constrain the masses and orbital elements using only
the transit times.  Naturally, this requires full dynamical integrations, in
which the planets cannot remain on circular orbits.  However, 
for planet b we assumed a circular orbit at the dynamical epoch,
since we are not attempting to fit its transit time variations.
All of the other orbital
parameters and masses were free to vary.  The resulting $\chi^2$ is 88.5 for
101 degrees of freedom (114 transit times, minus 5 parameters for planets c
and d, minus 3 for \koithree), which is quite acceptable.  To be compared with
the radial-velocity solution, the solved-for masses were [$m_b, m_c, m_d$] =
[$18 \pm 9$, $17.3 \pm 1.7$, $15.8 \pm 1.3$]\mearth, shown in the third row of
Table~\ref{tab:FinalMasses}.   These solutions had extremely low
eccentricities ($e<0.003$) for {\planetc} and {\planetd}.  As above, the
innermost planet is only very marginally detected.  In contrast to the
radial-velocity solution, however, the masses of the interacting planets are
very precisely pinned down by the large variations seen in
Figure~\ref{fig:omc}.

Finally, we generated a joint solution to the transit times and radial
velocities.  Graphically, this solution is given by Figure~\ref{fig:omc}.
A $\chi^2=103.4$ for 114 degrees of freedom is achieved, an excellent fit to
the data.  The orbital parameters and their formal errors (the output of
the Levenberg-Marquardt algorithm) are given in Table~\ref{tab:TTVsolution},
and the planetary masses are given in the fourth row of
Table~\ref{tab:FinalMasses}.

In the previous subsection we found no drift in the inclinations of the
three planets. We can use the $3 \sigma$ upper limits on
$| \Delta i/\Delta E |$
of [$37$, $70$, $33$]$\times 10^{-4}$ degrees per epoch for
{\planetb}, c, and d to place limits on their mutual inclination
\citep{ME02}.  To do this, we measure the value of
$| \Delta i/\Delta E |$
seen in numerical simulation, which depends nearly linearly on the
difference in nodal angle on the sky of two planets \citep{BaChCh10}.  Taking
the masses and orbits from the best-fit TTV/RV solution above, we simulated
{\planetc} and d with 1 degree (and 10 degrees) of mutual inclination, we
find $| \Delta i/\Delta E |_c = 1.8\times10^{-4}$ degrees per epoch
($16\times10^{-4}$ degrees per epoch) and
$| \Delta i/\Delta E |_d = 2.9\times10^{-4}$ degrees per epoch
($26\times10^{-4}$ degrees per epoch).  The interaction between {\planetb}
and c is also potentially observable; simulating their orbits with 1 (10)
degree(s) of mutual inclination, we find
$| \Delta i/\Delta E |_b$=$1.3 (11) \times10^{-4}$ degrees per epoch and
$| \Delta i/\Delta E |_c$=$0.82 (7.4) \times10^{-4}$ degrees per epoch.
The interaction between {\planetb} and d is considerably weaker,
with a drift of $\sim2 \times 10^{-4}$ degrees per epoch even for
10 degree mutual inclination,
therefore we ignore it when setting mutual inclination limits.
To find the $3 \sigma$ upper limit to the nodal difference of c and d,
we compare the 10-degree calculated value of
$| \Delta i/\Delta E |_d$ to its observational $3 \sigma$
upper limit, so we find $| \Omega_c - \Omega_d | < 13^\circ$.
The inclination difference in the complementary direction is only
$i_d-i_c = 0.40 \pm 0.24$~degrees, so the limit on the true mutual
inclination is $i_{cd}<13$~degrees.
The nodal difference of b and c is more poorly constrained,
as the calculated value of $| \Delta i/\Delta E |$ for both planets in the
10-degree mutual inclination simulation is small compared to its
observational $3 \sigma$ upper limit.  Thus a moderate ($\sim 20^\circ$)
mutual inclination is permissible, which would be large enough to affect the
timing fits. In that case, the model should self-consistently fit the
radial velocities, transit times, \emph{and} transit durations. 
However, we defer dynamical interpretation of the mutual inclination
of the inner planet to the others until more data are gathered,
which should tighten the limit.

%
%

%
%

%
%

\begin{deluxetable*}{ccccc}
\tablenum{7}
\tablewidth{0pt}
\tablecaption{Osculating Jacobian Elements, at epoch 2455168.0 [BJD], for the 3-planet TTV Dynamical Solution. \label{tab:TTVsolution}}
\tablehead{
\colhead{} &
\colhead{Period (days)} &
\colhead{T$_0$ (days)}  &
\colhead{$e \cos \omega$ } &
\colhead{$e \sin \omega$ }
}
\startdata
b &  $3.504674 \pm 0.000054$ & $266.276996 \pm 0.005453$ & 0 & 0 \\
c &  $7.641039 \pm 0.000087$ & $267.092502 \pm 0.000262$ & $0.000291 \pm 0.000079$ & $0.000173 \pm  0.000233$ \\
d & $14.860509 \pm 0.000148$ & $269.174850 \pm 0.000253$ & $ -0.000076 \pm 0.000019$ & $0.000516 \pm 0.000450$ \\
\enddata
\end{deluxetable*}

\begin{deluxetable*}{lcccc}
\tablenum{8}
\tablewidth{0pt}
\tablecaption{Masses and Densities of the Planets in the {\starname} System
\label{tab:FinalMasses}}
\tablehead{\colhead{Method} & & \colhead{\planetb} &
\colhead{\planetc} & \colhead{\planetd} \\
&&\colhead{(\koithree)}&\colhead{(\koione)}&\colhead{(\koitwo)}}
\startdata
MCMC (lightcurve + RV) Solution & (\mearth) & \MCMCmplanetb & \MCMCmplanetc & \MCMCmplanetd \\
RV + transit time $P$ and T$_0$ & (\mearth) & $12 \pm 5$ & $15 \pm 5$ & $28 \pm 7$ \\
TTV dynamical model & (\mearth) & $18 \pm 9$ & $17.3 \pm 1.7$ & $15.8 \pm 1.3$ \\
TTV + RV dynamical model (adopted values) & (\mearth) & \TTVmplanetb &
\TTVmplanetc & \TTVmplanetd \\
\hline
density from adopted mass & (${\rm g~cm}^{-3}$) & \TTVrhoplanetb & \TTVrhoplanetc & \TTVrhoplanetd \\
\enddata
\end{deluxetable*}

%
%

\section{\blender\ Analysis of {\koithree}}
\label{sec:blender}

The lack of a clear dynamical confirmation of the nature of {\koithree}
requires us to examine the wide variety of astrophysical false
positives (blends) that might mimic the photometric transit, and to
assess their {\it a priori\/} likelihood compared to that of a true
planet. For this we apply the \blender\ technique described by
\cite{ToKoSa04,ToFrBa11}, with further developments as reported by
\cite{FrToDe11}.  \blender\ uses the detailed shape of the transit
light curve to weed out scenarios that lead to the wrong shape for a
transit.  The kinds of false positives we are concerned with here
include background or foreground eclipsing binaries blended with the
target, as well as physically associated eclipsing binaries, which
generally cannot be resolved in high-angular resolution imaging. In
each case the pair of eclipsing objects can also be a star transited
by a larger planet. Briefly, \blender\ simulates a very large number
of light curves resulting from these blend scenarios with a range of
stellar (or planetary) parameters, and compares them to the {\kepler}
photometry in a $\chi^2$ sense. Blends providing poor fits are
considered to be ruled out, enabling us to place constraints on the
kinds of objects composing the eclipsing pair that yield viable
blends, including their size or mass, as well as other properties of
the blend such as the overall brightness and color, and even the
eccentricities ($e$) of the orbits.  We refer the reader to the above
references for details.  Following the nomenclature in those sources,
the objects in the eclipsing pair are designated the `secondary' and
`tertiary', and the target itself is the `primary'. Stellar properties
are drawn from model isochrones.

\begin{figure}
\plotone{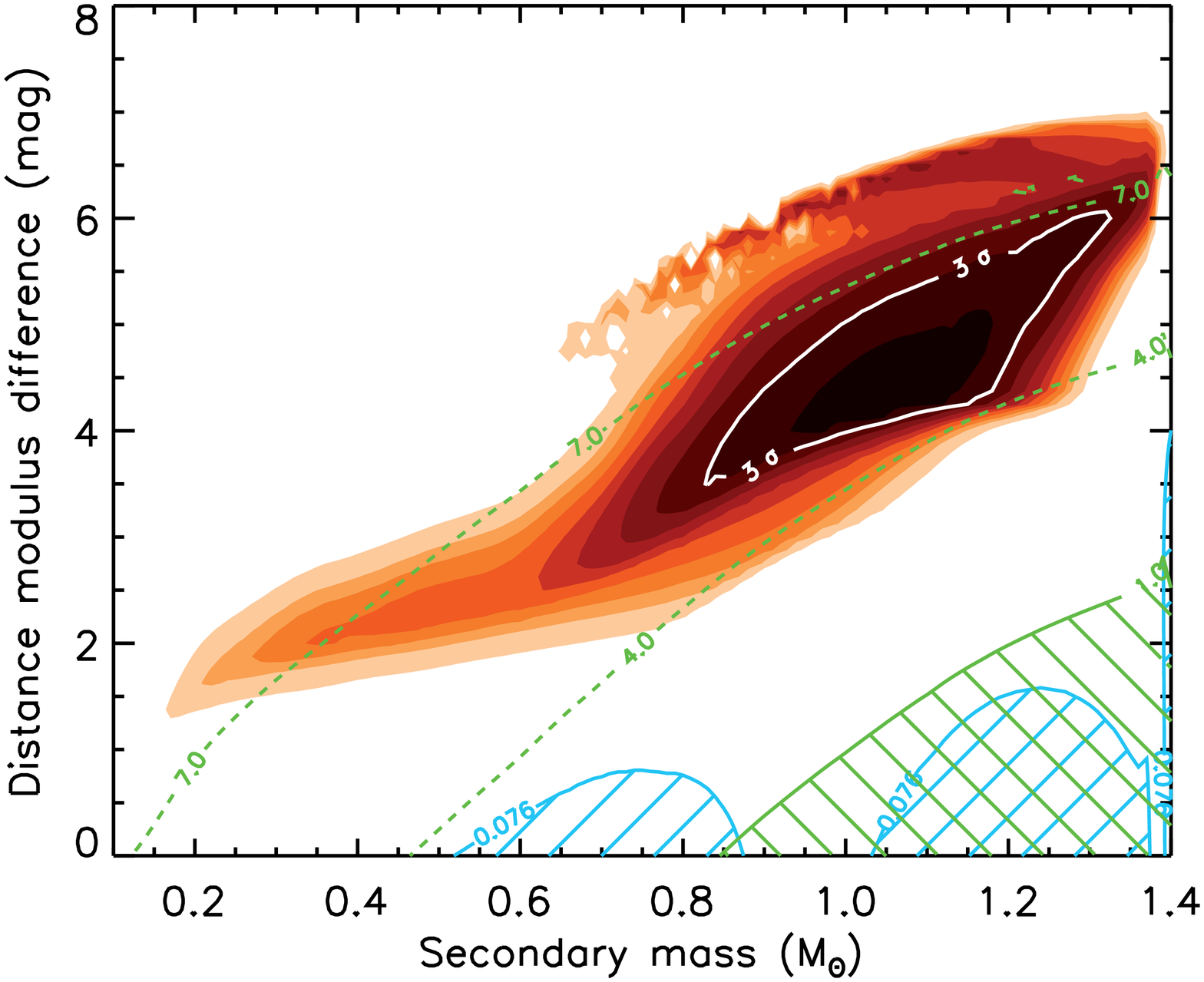}
\caption{Map of the {\planetb} ({\koithree})
$\chi^2$ surface (goodness of fit) for blends
involving background eclipsing binaries with stellar tertiaries and
arbitrary orbital eccentricities.  The vertical axis represents the
linear separation between the background binary and the primary, cast
for convenience in terms of the difference in the distance
modulus. Only blends within the solid white contour match the {\kepler}
light curve within acceptable limits \citep[3$\sigma$, where $\sigma$
is the significance level of the $\chi^2$ difference compared to a
transiting planet model; see][]{FrToDe11}. Other concentric colored
areas represent increasingly worse fits (4$\sigma$, 5$\sigma$, etc.),
and correspond to blends we consider to be ruled out. Dashed green
lines are labeled with the magnitude difference $\Delta K\!p$ between
the blended binary and the primary, and encompass the brightness range
allowed by \blender\ ($4 \leq \Delta K\!p \leq 7$). Blends with
eclipsing binaries bright enough to be detected spectroscopically
($\Delta K\!p \leq 1$~mag) are indicated with the hatched region below
the solid green line, but are already ruled out by \blender. Similarly
with the blue hatched areas that mark blends that are either too red
or too blue compared to the measured color (see text and
Figure~\ref{fig:bp}). When further constraining these blends to have
realistic eccentricities ($e \leq 0.1$; see text), we find that
\emph{all} of them are excluded by \blender.\label{fig:bs}}
\end{figure}

Simulations with \blender\ indicate that background eclipsing binaries
with two stellar components can only produce viable false positives if
they are restricted to a narrow range of masses for the secondaries
(approximately $0.8 \leq M_2/\msun \leq 1.3$), as well as a limited
interval in brightness ($K\!p$ magnitude) relative to the target ($4
\leq \Delta K\!p \leq 7$).  This is illustrated in
Figure~\ref{fig:bs}, in which we show the $\chi^2$ landscape from
\blender\ for all blend fits of this kind. Regions outside the
3-$\sigma$ contour correspond to scenarios with transit shapes that do
not provide acceptable fits to the {\kepler} photometry, i.e., fits
that are much worse than a true planet fit. These configurations are
therefore excluded.

%
%

\begin{figure}
\plotone{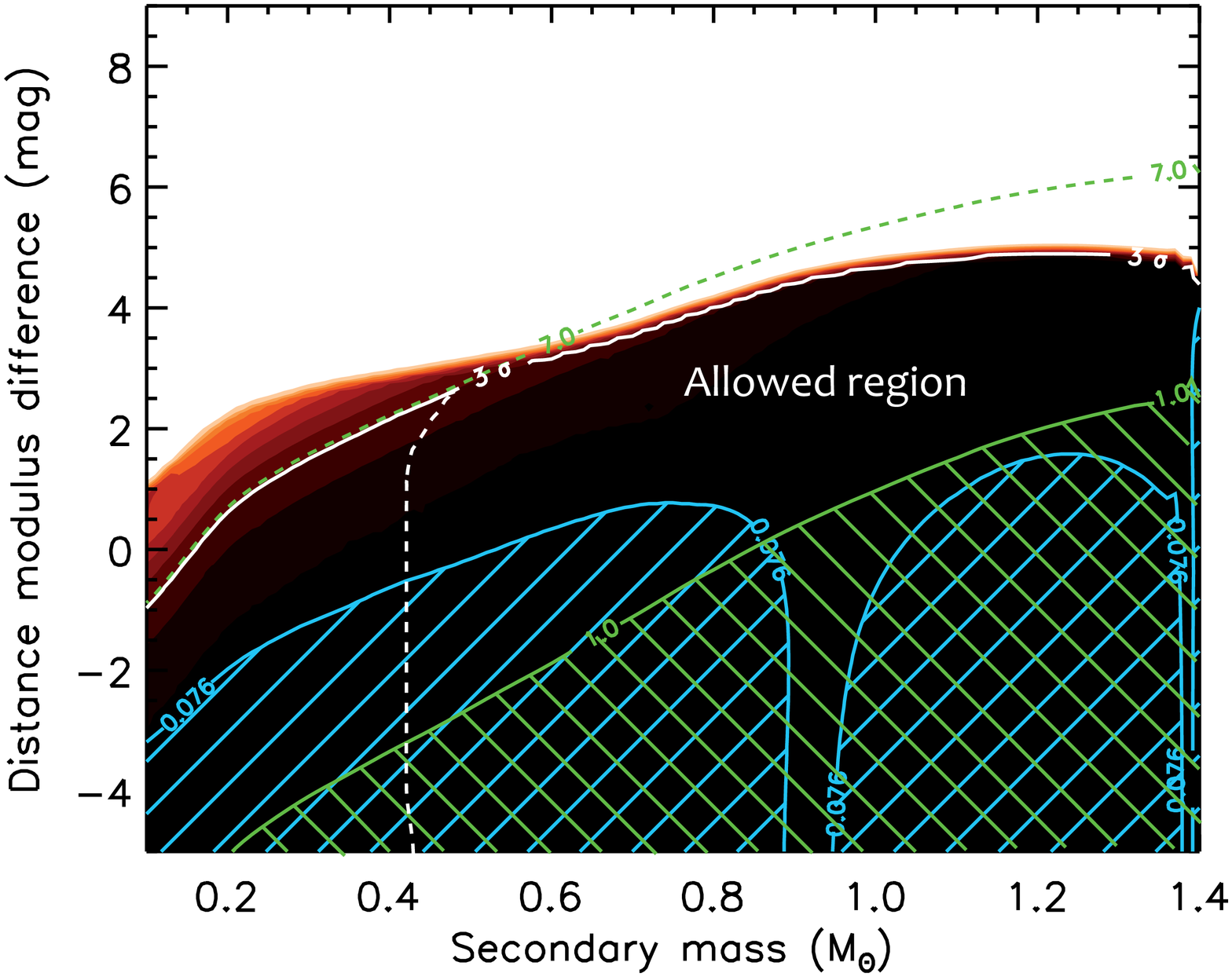}
%
\caption{Similar to Figure~\ref{fig:bs} (and with the same color
scheme) for {\planetb} ({\koithree})
blends involving background or foreground stars transited
by a larger planet. The blue hatched regions correspond to blends that
are too red (left) or too blue (right) compared to the measured
$r-K_s$ color of {\starname}, and are thus ruled out. When blends are
restricted to realistic orbital eccentricities ($e \leq 0.3$), many of
the later-type secondaries are excluded (dashed 3$\sigma$ contour) The
combination of the brightness (green hatched area) and color
constraints leaves only a reduced area of parameter space (``Allowed
region'') where blends are a suitable alternative to a transiting
planet model. All of these scenarios have $\Delta K\!p < 7.0$ (dashed
green line).\label{fig:bp}}
\end{figure}

For blends involving a background/foreground star transited by a
larger planet, there is in principle a wide range of allowed masses
(spectral types) for the secondary stars, as shown in
Figure~\ref{fig:bp}. However, other constraints available for
{\starname} strongly limit the number of these false positives. In
particular, by comparing the predicted $r-K_s$ color of each blend
against the measured color of the star from the KIC \citep[$r-K_s =
1.723 \pm 0.031$;][]{BrLaEv11}, we find that many of the smaller-mass
secondaries are ruled out because the blends would be much too red
compared to the known color index of {\starname}  (by more than
3$\sigma$). Others are excluded because the secondary star would be
very bright (within one magnitude of the target, or in some cases even
brighter than the target), and would have been noticed
spectroscopically, if unresolved in our AO or speckle imaging. This
removes many but not all blends of this kind.

%
%

\begin{figure}
\plotone{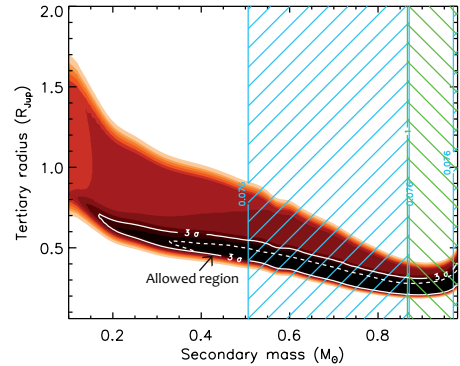}
%
\caption{Similar to Figure~\ref{fig:bs} for {\planetb} ({\koithree})
for the case of hierarchical
triple systems in which the secondary is transited by a planet. After
taking into account the constraints on the $r-K_s$ color and
brightness (blue and green hatched regions, respectively), only
secondary stars with $M_2 \leq 0.5$\,$\msun$ lead to blend light
curves that match the observations. Further restriction to realistic
orbital eccentricities ($e \leq 0.3$; see text) leads to a slightly
smaller 3-$\sigma$ contour (dashed). \label{fig:htp}}
\end{figure}

For eclipsing binaries that are physically associated with the target
(in a hierarchical triple star configuration) we find that the blend
light curves invariably have the wrong shape to mimic a true
transiting planet signal, for any combination of stellar parameters
for the secondary and tertiary. Either the depth, duration, or
steepness of the ingress/egress phases of the transits provide a poor
match to the {\kepler} photometry.  These scenarios are therefore all
excluded. On the other hand, if we allow the tertiaries to be planets,
then we do find a variety of secondary masses that can produce viable
blends when transited by a planet of the appropriate size. The
$\chi^2$ map for this general case appears in Figure~\ref{fig:htp},
and shows the range of radii permitted for the tertiaries, as well as
the interval of secondary masses that yield suitable blends.

%
%

The duration of a transit is set by the length of the chord traversed
by the tertiary and the tangential velocity of the tertiary during the
event. The chord length, in turn, depends on the size of the secondary
and the impact parameter. Therefore, the measured duration of a
transit provides a strong constraint on the allowed sizes for the
secondary stars (or equivalently, their masses or spectral types).  In
the above \blender\ simulations we have placed no restriction on the
orbital eccentricities of the star+star or star+planet pairs that can
be blended with the target. When the orbits are permitted to be
eccentric, the tangential velocity of the tertiary during transit can
be significantly different from the circular case, and this allows a
much larger range of secondary sizes than would otherwise be possible.
In particular, chance alignments with later-type stars transited by a
planet near apoastron become viable as blends, and represent a good
fraction of the false positives shown in Figures~\ref{fig:bp} and
\ref{fig:htp}. Some blends involving larger secondaries transited at
periastron can also provide acceptable fits.

We note, however, that the period of {\koithree} is relatively short (3.5
days), and very large eccentricities, such as some of our simulated
blends have, are unlikely as they would be expected to be damped by
tidal forces \citep[see][]{Ma08}. Indeed, among binaries with
main-sequence primary stars that are similar to {\starname}  (solar-type,
or later), all systems under 3.5 days have essentially circular orbits
\cite[see, e.g.,][]{HaMaUd03,RaMAHe10}. We may take $e = 0.1$
as a conservative upper limit.  Similarly, among the known transiting
planets with periods of 3.5 days or shorter and host stars of any
spectral type, none are found to have eccentricities as large as $e =
0.3$.  When constraining the false positives for {\koithree} to be within
these eccentricity limits, we find that \emph{all} background
eclipsing binaries (stellar tertiaries) are easily excluded as they
all require fairly eccentric orbits in order to match the observed
duration. Additionally, the numbers of blends involving star+planet
pairs in the foreground/background or in hierarchical triple systems
are considerably reduced when restricting the eccentricities, although
many remain that we can not rule out. In the following we assess their
frequency, and compare it with the expected frequency of transiting
planets.

\subsection{Validating {\koithree}}\label{sec:validation}

The {\it a priori\/} frequency of stars in the background or
foreground of the target that are orbited by a transiting planet and
are capable of mimicking the {\koithree} signal may be estimated from the
number density of stars in the vicinity of {\starname}, the area around
the target in which such stars would go undetected in our
high-resolution imaging, and the frequency of transiting planets with
the appropriate characteristics.  To obtain the number density (stars
per square degree) we make use of the Galactic structure models of
\cite{RoReDe03}, and we perform this calculation in half-magnitude
bins, as shown in Table~\ref{tab:blendfreq}. For each bin we further
restrict the star counts using the constraints on the mass of the
secondaries supplied by \blender\ (see Figure~\ref{fig:bp}), and the
eccentricity limit for transiting planets discussed above ($e \leq
0.3$). These mass ranges are listed in column~3, and the resulting
densities appear in column~4. Bins with no entries correspond to
brightness ranges excluded by \blender. The maximum angular separation
($\rho_{\rm max}$) at which stars of each brightness would escape
detection in our AO/speckle imaging is shown in column~5 (see
Table~\ref{tab:paloAOlimits} and Section~\ref{sec:imaging}).
The result for the number of stars in each magnitude bin is given in
column~6, in units of $10^{-6}$.

%
%

\begin{deluxetable*}{ccccccc}
\tablenum{9}
\tabletypesize{\scriptsize}
\tablewidth{0pc}
\tablecaption{Blend frequency estimate for {\koithree}. \label{tab:blendfreq}}
\tablehead{
& &
\multicolumn{5}{c}{Blends Involving Planetary Tertiaries} \\[+1.5ex]
\cline{3-7} \\ [-1.5ex]
\colhead{$K\!p$ Range} &
\colhead{$\Delta K\!p$} &
\colhead{Stellar} &
\colhead{Stellar Density} &
\colhead{$\rho_{\rm max}$} &
\colhead{Stars} &
\colhead{Transiting Planets} \\
\colhead{(mag)} &
\colhead{(mag)} &
\colhead{Mass Range} &
\colhead{(per sq.\ deg)} &
\colhead{(\arcsec)} &
\colhead{($\times 10^{-6}$)} &
\colhead{0.32--1.96\,$R_{\rm Jup}$, $f_{\rm planet}=0.24$\%} \\
\colhead{} &
\colhead{} &
\colhead{($M_{\odot}$)} &
\colhead{} &
\colhead{} &
\colhead{} &
\colhead{($\times 10^{-6}$)} \\
\colhead{(1)} &
\colhead{(2)} &
\colhead{(3)} &
\colhead{(4)} &
\colhead{(5)} &
\colhead{(6)} &
\colhead{(7)}
}
\startdata
13.5--14.0  &  0.5 & \nodata   & \nodata&\nodata & \nodata& \nodata \\
14.0--14.5  &  1.0 & \nodata   & \nodata&\nodata & \nodata& \nodata \\
14.5--15.0  &  1.5 & 0.87--1.40  & 862    &  0.08  &  1.34   & 0.003 \\
15.0--15.5  &  2.0 & 0.82--1.40  & 1377   &  0.11  &  4.04   & 0.010 \\
15.5--16.0  &  2.5 & 0.78--1.40  & 2094   &  0.13  &  8.58   & 0.021  \\
16.0--16.5  &  3.0 & 0.72--1.40  & 3053   &  0.16  &  18.9   & 0.045  \\
16.5--17.0  &  3.5 & 0.43--1.40  & 4341   &  0.20  &  42.1   & 0.101  \\
17.0--17.5  &  4.0 & 0.43--1.40  & 5873   &  0.24  &  82.0   & 0.197  \\
17.5--18.0  &  4.5 & 0.43--1.32  & 7599   &  0.32  &  189    & 0.454 \\
18.0--18.5  &  5.0 & 0.43--1.25  & 9399   &  0.40  &  365    & 0.876 \\
18.5--19.0  &  5.5 & 0.43--1.09  & 10819  &  0.56  &  822    & 1.973 \\
19.0--19.5  &  6.0 & 0.43--1.01  & 11988  &  0.64  &  1190   & 2.856 \\
19.5--20.0  &  6.5 & 0.43--0.92  & 12585  &  0.80  &  1952   & 4.685 \\
20.0--20.5  &  7.0 & 0.43--0.61  & 3373   &  0.88  &  633    & 1.519 \\
20.5--21.0  &  7.5 & \nodata     & \nodata&\nodata & \nodata& \nodata \\
21.0--21.5  &  8.0 & \nodata     & \nodata&\nodata & \nodata& \nodata \\
\noalign{\vskip 6pt}
\multicolumn{2}{c}{Totals} & & & & 5308 & 12.7 \\
\noalign{\vskip 4pt}
\hline
\noalign{\vskip 4pt}
\multicolumn{7}{c}{Blend frequency from hierarchical triples (see text) = $4.4 \times 10^{-6}$} \\
\noalign{\vskip 4pt}
\hline
\noalign{\vskip 4pt}
\multicolumn{7}{c}{Total frequency (BF) = $(12.7 + 4.4) \times 10^{-6} = 17.1 \times 10^{-6}$} \\
\enddata
\tablecomments{Magnitude bins with no entries correspond to brightness
ranges in which \blender\ excludes all blends.}
\end{deluxetable*}

To estimate the frequency of transiting planets that might be expected
to orbit these stars (and lead to a false positive) we rely on the
results from \citet{BoKoBa11b}, who reported a total of 1235 planet
candidates among the 156,453 {\kepler} targets observed during the
first four months of the Mission.  These signals have not yet been
confirmed to be caused by planets, and therefore remain candidates
until they can be thoroughly followed up. However, the rate of false
positives in this sample is expected to be quite small
\citep[$\sim$10\% or less; see][]{MoJo11}, so our results will not
be significantly affected by the assumption that all of the candidates
are planets. We further assume that the census of \cite{BoKoBa11b}
is largely complete. After accounting for the additional \blender\
constraint on the range of planet sizes for blends of this kind
(tertiaries of 0.32--1.96\,$R_{\rm Jup}$), we find that the transiting
planet frequency is $f_{\rm planet} = 374/156,\!453 = 0.0024$.
Multiplying this frequency by the star counts in column~6 of
Table~\ref{tab:blendfreq} we arrive at the blend frequencies listed in
column~7, which are added up in the ``Totals'' line of the table
($12.7 \times 10^{-6}$).

Next we address the frequency of hierarchical triples, that is,
physically associated companions to the target that are orbited by a
larger transiting planet able to mimic the signal. The rate of
occurrence of this kind of false positive may be estimated by
considering the overall frequency of binary stars \citep[34\%
according to][]{RaMAHe10} along with constraints on the mass range
of such companions and how often they would be orbited by a transiting
planet of the right size (0.43--0.53\,$R_{\rm Jup}$; see
Figure~\ref{fig:htp}).  The mass constraints include not only those
coming directly from \blender, but also take into consideration the
color and brightness limits mentioned earlier.  We performed this
calculation in a Monte Carlo fashion, drawing the secondary stars from
the mass ratio distribution reported by \cite{RaMAHe10}, and the
transiting planet eccentricities from the actual distribution of known
transiting planets (http://exoplanet.eu/), with repetition. Planet
frequencies in the appropriate radius range were taken as before from
\cite{BoKoBa11b}. The result is a frequency of hierarchical triples
of $4.4 \times 10^{-6}$, which we list at the bottom of
Table~\ref{tab:blendfreq}.

Combining this estimate with that of background/foreground star+planet
pairs described previously, we arrive at a total blend frequency of
${\rm BF} = (12.7 + 4.4) \times 10^{-6} \approx 1.7 \times 10^{-5}$,
which represents the {\it a priori\/} likelihood of a false positive.
From a Bayesian point of view analogous to that adopted to validate
previous {\kepler} candidates, our confidence in the planetary nature
of {\koithree} will depend on how this likelihood compares to the {\it a
priori\/} likelihood of a true transiting planet, addressed below.

The blend frequency of $1.7 \times 10^{-5}$ corresponds to false
positive scenarios giving fits to the {\kepler} photometry that are
within 3$\sigma$ of the best planet fit. We use a similar criterion to
estimate the {\it a priori\/} transiting planet frequency by counting
the KOIs in the \cite{BoKoBa11b} sample that have radii within
3$\sigma$ of the value determined from the best fit to {\koithree} ($R_p =
\rplanetb\,$\rearth).  We find 284 that are within this
range, giving a planet frequency ${\rm PF} = 284/156,\!453 = 1.8
\times 10^{-3}$.

This estimate does not account for the fact that the geometric transit
probability of a planet is significantly increased by the presence of
additional planets in the system ({\planetc} and {\planetd} in
this case), given that mutual inclination angles in systems with
multiple transiting planets have been found be relatively small
\citep[typically 1--4\arcdeg;][]{LiRaFa11}. Furthermore, a planet
with the period of {\koithree} would be \emph{interior} to the other two,
further boosting the chances that it would transit.  To incorporate
this coplanarity effect, we have developed a Monte Carlo approach,
described fully in Appendix~\ref{ap:mutualinc}, in which we
simulate randomly distributed reference planes and inclination
dispersions around this plane, from which a weighted distribution of
the inclination with respect to the line of sight for a third planet
is calculated.  Inclination angles relative to the random reference
plane are assumed to follow a Rayleigh distribution
\citep[see][]{LiRaFa11}. Although the known planets carry some
information on the inclination dispersion, the probability of transit
still depends somewhat on the allowed range of dispersion widths. When
the assumed prior for the inclination dispersion is uniform up to
4\arcdeg, following \cite{LiRaFa11}, we find that the flatness of
the system results in a very significant increase in the transit
probability for {\koithree} from 11.7\% to 97\%. To be conservative, we
adopt a larger range of possible inclination dispersions from
0\arcdeg\ to 10\arcdeg, motivated by the upper limit from the similar
Kepler-9 system \citep{HoFaRa10}. With this prior, the transit
probability for {\koithree} becomes 84\%, or an increase by a factor of
$\sim$7 over the case of a single transiting planet.

Thus, the likelihood of a planet is more than 700 times greater (${\rm
PF/BF} = 0.013/1.7\times10^{-5} \approx 700$) than that of a false
positive, which we consider sufficient to validate {\koithree} as a true
planet with a high degree of confidence. We designate this planet
\planetb.  We note that our planet frequency calculation assumes
the 1235 candidates cataloged by \cite{BoKoBa11b} are all true
planets.  If we were to suppose conservatively that as many as 50\%
are false positives \citep[an unlikely proposition that is also
inconsistent with other evidence; see][]{BoKoBa11b,HoMaJo10}, the
planet likelihood would still be $\sim$350 times greater than the
likelihood of a blend, implying a false alarm rate sufficiently small
to validate the candidate.

\section{Physical Properties of the Planets}\label{sec:discussion}

The {\starname} system consists of two low-density Neptune-mass planets
near a 2:1 mean motion resonance and an inner super-Earth-size planet.
Its architecture bears a strong resemblance to Kepler-9,
except that the {\starname} system is less compact and its planets
are less dense.
The use of the observed transit times as well as the radial velocity data
in the dynamical model of the {\starname} system places tight limits on
the allowed planetary masses.  
We adopt these values as our best determination of the masses of the
transiting planets in the {\starname} system.
The last line in Table~\ref{tab:FinalMasses} gives the planet densities,
computed from the final adopted planet masses.
The TTV measurements together with the radial velocities restrict the
masses of {\planetc} (\TTVmplanetc \mearth)
and {\planetd} (\TTVmplanetd \mearth) to be similar to each
other.  Both are slightly lower than the mass of Neptune.
Their radii however are 40\% and 80\% larger than Neptune respectively,
giving them bulk densities of \TTVrhoplanetc\,${\rm g~cm}^{-3}$ and
\TTVrhoplanetd\,${\rm g~cm}^{-3}$, which are only 0.36 and 0.16 that of Neptune.
The mass of {\planetb} from the joint dynamical solution to the transit
times and the RV measurements, is \TTVmplanetb \mearth.  With its
``super-Earth'' size radius of \rplanetb\,{\rearth}, the density of this inner
planet in the system is \TTVrhoplanetb\,${\rm g~cm}^{-3}$.

Using the methods described in \citet{MiFoJa09} and \citet{MiFo11}, we have
modeled the thermal evolution and interior structure of the two
``Neptune-class'' planets, {\planetc} and {\planetd}.
Both planets have inflated radii compared to Uranus and Neptune,
which points to two effects. 
The first is that the high incident flux slows their contraction. 
The second is that the mass fraction of heavy element within these
two planets is lower than that of Uranus and Neptune, meaning the
mass fraction of H-He gas is larger. 
Uranus and Neptune are $\sim$ 80-90\% heavy elements (e.g., water
and rock) by mass \citep{FoNe10}, while below we show that the heavy element
mass fractions of {\planetc} and {\planetd} are somewhat lower than
these values.

%
%

\begin{figure*}
\plotone{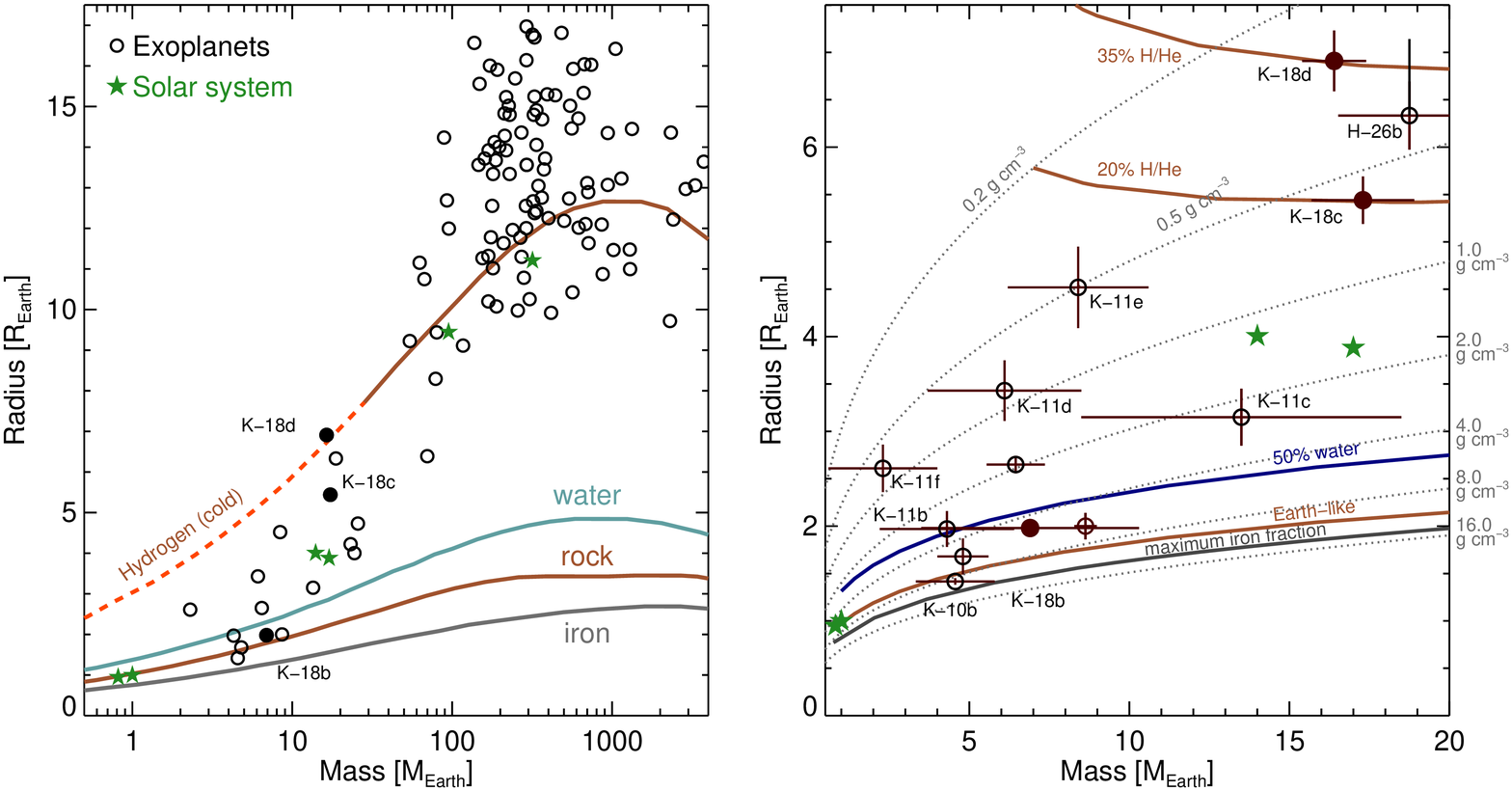}
\caption{The mass-radius diagram for transiting exoplanets (circles) and
Solar System planets (stars). Left panel - Known transiting exoplanets (open
circles) and {\planetb}, c and d (filled circles), with curves of theoretical
relations for cold pure hydrogen, water, rock (silicate), and iron (after
\citet{SeKuHM07}; \citet{FoMaBa07}).  Two detailed models with mixtures
and surface temperatures appropriate for {\planetb} and {\planetc}
are also shown \citep{MiFo11}.
Right panel - Zoom on the smaller planets, with curves of constant mean
density and curves of detailed interior models of constant composition. The
theoretical models are (from top to bottom): ``Neptune-class'' models with
hydrogen/helium envelopes and 50-50 ice-rock (by mass) cores from thermal
evolution calculations \citep{MiFo11} - the labeled H/He fractions
are averages - if all metals are in a core, these fractions will be 31\%
and 16\%, respectively; if mixed partially, they become 38\% and 23\%,
respectively, for the curves shown; the `50\% water' models have
compositions of 44\% silicate mantle and 6\% iron core, and the nominal
`Earth-like' composition with terrestrial iron/silicon ratio and no
volatiles are by \citet{VaOCSa06} and Zeng \& Sasselov (2011, subm.) The
maximum mantle stripping limit (maximum iron fraction, minimum radius) was
computed by \citet{MaSaHe10}.  All these model curves of constant
composition are for illustration purposes, as the degeneracy between
composition mixtures and mean density is significant in this part of the
mass-radius diagram.
The data for the exoplanets were taken from \citet{QuBoMo09}, \citet{ChBeIr09},
\citet{HaBaKi11}, \citet{BaBoBr11}, \citet{LiFaFo11} and \citet{WiMaDa11}. 
We note that new analysis of CoRoT-7b \citep{HaFrNa11}
places it at a similar mass to \planetb, and similar
high-density composition to Kepler-10b. The three unmarked exoplanets
surrounding {\planetb} in the diagram are (in increasing mass) CoRoT-7\,b,
GJ\,1214\,b, and 55\,Cnc\,e.
\label{fig:mass_radius}}
\end{figure*}

Thermal evolution/contraction models are constrained such that the radius
of each planet must be reproduced at the system's estimated age.  As in
\citet{MiFo11}, all relevant uncertainties are accounted for.  These
include uncertainties in the age of the system, the semimajor axes, masses,
and radii of the planets, and the distribution of the heavy elements within
each planet.  We do not include an internal heating contribution due
to eccentricity damping in either planet, as this power source is
expected to fall off as $a^{-15/2}$ \citep{JaGrBa08} and, furthermore,
the eccentricities suggested by the TTV solutions are quite small.
A 50-50 by mass ice-rock equation of state is used for the
heavy elements.  We find a heavy element mass of $13.5 \pm 1.8${\mearth}
in {\planetc} ($\sim$~80\% of the planet's mass) and $10.1 \pm
1.4${\mearth} in {\planetd} ($\sim$~60\% of the planet's mass).  {\planetc} is
clearly more ``metal-rich'' than {\planetd}.  Both planets are consistent with
a core-accretion formation scenario in which $\sim$~10{\mearth} of
heavy elements gravitationally captures an envelope of H-He gas.  This
envelope itself may be enhanced in heavy elements, as is inferred for Uranus
and Neptune.

Planet {\planetd}, with a large radius of nearly 7{\rearth}, may point to a
population of Neptune-mass exoplanets having relatively low heavy element
mass fractions and radii approaching that of the gas giant regime. They
appear very similar to HAT-P-26b, a 4.2 day planet orbiting a cooler
K1-dwarf. The formation and evolution of such lower-density Neptune-class
planets was recently studied in detail by \citet{RoBoLi11}. They find
that modestly more massive H-He envelopes than found on Uranus and Neptune
(leading to larger planetary radii and lower densities) may be a common
outcome of the core-accretion planet formation process.

The inner, 3.5-day period planet \planetb, is a super-Earth 
that requires a dominant mixture of water ice and rock, and no
hydrogen/helium envelope. While the latter cannot be excluded simply on the
basis of the planet's mass and radius, the evaporation timescale for a
primordial H/He envelope for a hot planet such as {\planetb} is much
shorter than the old age derived for the {\starname} system, and such
a H/He envelope should not be present.
Thus, despite its lower equilibrium temperature, {\planetb} resembles
55\,Cnc\,e and CoRoT-7b (as originally measured by \citet{QuBoMo09}
though the \citet{HaFrNa11} re-analysis makes CoRoT-7b very similar to
Kepler-10b).  \planetb, together with 55\,Cnc\,e \citep{WiMaDa11},
are likely our best known cases yet of water planets with
substantial steam atmospheres (given their high surface temperatures).

It is interesting to compare the three transiting planets in {\starname} in
terms of their apparent compositions and orbital sequence. {\starname}
reinforces a pattern already seen in Kepler-11, and with less confidence in
Kepler-10 and Kepler-9. Namely, inner planets are denser, though not always
by very much - compare {\planetb} vs. c \& d, and Kepler-11b,c vs. d, e, \& f
(Figure 12). It remains unclear at present whether this reflects a density
gradient at formation or could be accomplished by evaporation later.

\acknowledgements
{\kepler} was competitively selected as the tenth Discovery mission.
Funding for the {\kepler} Mission is provided by NASA's Science Mission
Directorate.   We are deeply grateful for the very hard work of the
entire {\kepler} team.
This research is based in part on observations made with the {\emph {Spitzer
Space Telescope}}, which is operated by the Jet Propulsion Laboratory,
California Institute of Technology under a contract with NASA.
Support for this work was provided by NASA through an award
issued by JPL/Caltech.
Some of the data presented herein were obtained at the
W.~M. Keck Observatory, which is operated as a scientific
partnership among the California Institute of Technology,
the University of California, and the National Aeronautics and
Space Administration. The Keck Observatory was made possible by the
generous financial support of the W. M. Keck Foundation.

\newcommand{\noopsort}[1]{}
  \newcommand{\printfirst}[2]{#1} \newcommand{\singleletter}[1]{#1}
  \newcommand{\switchargs}[2]{#2#1}

\appendix

\section{Incorporating Coplanarity in Multi-Transiting Systems to
Estimate Transit Probability}
\label{ap:mutualinc}

In a system with $N$ transiting planets, what is the geometric
probability that an additional planet (called ``planet $N+1$'') with a
given period would transit, taking into account the fact that
planetary orbits are expected (or observed) to be nearly coplanar? By
incorporating the effect of coplanarity, the probability is
significantly increased for additional planets to transit in known
transiting planet systems compared to isotropically distributed
planets.

In a one-planet, one-candidate system, $N=1$ and there is not much
that can be done unless a prior assumption for the true mutual
inclination is taken \citep{BeSe10}.  While in some cases,
the true mutual inclination can be directly measured (see \citet{RaHo10}
for various methods), generally it will have to be inferred from
population statistics.  \cite{LiRaFa11} estimate that the
inclination distribution of short-period planetary systems seen by
{\kepler} ranges from 1--4\arcdeg. This range could be used as a prior,
both in the case of $N=1$ and in the case of higher multiplicities.
Additionally, with careful analysis, the absence of transit timing and
duration variations can be used to put an upper limit on mutual
inclinations, as in the case of Kepler-9 where the mutual inclination
must be less than 10\arcdeg\ \citep{HoFaRa10}.

However, systems with two or more transiting planets, such as
\starname, have evidence of being thin without direct reference to the
greater population of {\kepler} multiples. Furthermore, the
inclinations of the known planets themselves give information on both
the location of the reference plane of that system (i.e., the Laplace
plane) and the typical inclination dispersion. If both planets have
the same inclination slightly different from 90\arcdeg, as in the case
of \planetc\ and d, then this suggests that the reference plane for
this system is something like the average plane between the two
planets and that the inclinations with respect to that plane are
likely quite small. In this situation, candidates with periods smaller
than the inner planet are quite likely to transit
\citep{RaHo10}.  If both planets have quite different
inclinations, as in the case of Kepler-10\,b and Kepler-10\,c
\citep{BaBoBr11}, this implies an uncertainty in the location of the
reference plane and a non-flat system, and the probability of
additional planets transiting is enhanced over completely isotropic
systems, but not as much \citep{FrToDe11}.

The goal is to quantify this effect in a natural way that uses all the
available information. This is done using a Monte Carlo simulation
that randomly generates reference planes, inclination dispersions, and
nodal angles. The Monte Carlo trials that result in systems that match
well the observed inclinations of the $N$ planets are given higher
weight than the vast majority of trials that do a very poor job. The
line-of-sight inclination of the additional planet is also calculated
in each Monte Carlo trial, allowing for the generation of a final
weighted distribution. The assumptions made in this calculation are
that exoplanet reference planes are, {\it a priori}, randomly oriented
in an isotropic way and that the inclinations (with respect to the
reference plane) of all the planets in a system are faithfully
represented by a single Rayleigh distribution. The latter assumption
does not account for the possible anticorrelation between size and
inclination that can occur in dynamically thermalized systems, i.e.,
larger planets may be more well-aligned than smaller planets. With the
Monte Carlo nature of the calculation and the error bars on
inclinations that are typically produced by \kepler, the importance of
a possible anticorrelation is reduced.

In practical terms, the Monte Carlo simulation draws a random
inclination ($i_r$) for the reference plane (uniform in $\cos i_r$).
The width of the Rayleigh distribution $\sigma_i$ is drawn randomly
from a distribution that is assumed \emph{a priori}. For each Monte
Carlo trial, a random inclination is drawn from the Rayleigh
distribution as is a random nodal angle for each of the $N+1$ planets;
the nodal angles are chosen uniformly between 0 and 2$\pi$. Using the
method in eq.~2 of \cite{RaHo10}, the on-the-sky inclinations of
the $N$ planets are calculated. These are compared to the observed
inclinations by computing the standard Gaussian $z$-score, i.e., by
calculating the number of standard deviations away the trial value is
from the observed value.  (This could be modified if the inclinations
and errors of the known planets are either not known or are known not
to be Gaussian.)
The assigned weight for each planet is equal to the area under the Gaussian
that has more extreme scores than the trial value, i.e., with a z-score of
$z = \frac{i_{MC} - i_{obs}}{i_{err}}$, the weight is $w = 1.0 -
\textrm{erf} ( \frac{|z|}{\sqrt{2}} ) $, where $i_{MC}$ is the calculated
trial inclination, $i_{obs}$ is the observed inclination with error
$i_{err}$, and erf is the standard error function.
If the trial
value of the inclination is exactly the same as the observed
inclination, the weight is 1; if it is many standard deviations off,
then the weight is essentially 0.  The total weight for a Monte Carlo
trial is the product of these weights over all the known $N$ planets
based on their inclinations. As expected, the weight increases when
the reference plane is taken near the average of the known planetary
inclinations with a Rayleigh width similar to the standard deviation
between the known inclinations.

Each of the Monte Carlo trials also calculates an inclination for
planet $N+1$, based on the same reference inclination and $\sigma_i$.
Using the weights derived from the observed inclinations of the $N$
transiting planets, the weighted distribution of the inclination of
planet $N+1$ can be created. To calculate the probability that planet
$N+1$ would be found transiting, the sum of the weights for those
simulations that would have produced a transiting planet is divided
by the sum of the weights for the entire Monte Carlo simulation.

Despite the information on the inclination dispersion from the known
planets that is included in the weighting, the answer depends somewhat
on the prior assumption for $\sigma_i$. For example, when applied to
{\starname}  with {\planetb} as planet $N+1$, this technique predicts a
transit probability of 47\%, 84\%, and 97\% when the prior on
$\sigma_i$ is uniformly drawn between 0\arcdeg\ and 90\arcdeg,
0\arcdeg\ and 10\arcdeg, and 0\arcdeg\ and 4\arcdeg, respectively. In
every case, the near coplanarity and low impact parameter of the two
outer planets significantly increase the probability that {\planetb}
is transiting. Changing the prior on the inclination of the reference
plane does not affect the result.
This method provides a quantitative way to estimate the increase in
transit probability for planets in multi-transiting systems. When
applied in combination with \blender\ and other techniques, it will
allow multi-transiting systems to be validated more easily than
singly-transiting systems. Note that the method described here only
estimates the improvement in the probability of the planet hypothesis
due to geometric constraints, and does not include the additional
effect that planets tend to be found in multiple systems, as discussed
in \cite{LiRaFa11}.


\begin{thebibliography}{82}
\expandafter\ifx\csname natexlab\endcsname\relax\def\natexlab#1{#1}\fi

\bibitem[{Agol {et~al.}(2005)Agol, Steffen, Sari, \& Clarkson}]{AgStSa05}
Agol, E., Steffen, J., Sari, R., \& Clarkson, W. 2005, MNRAS, 359, 567

\bibitem[{Ballard {et~al.}(2010)Ballard, Christiansen, Charbonneau, Deming,
  Holman, Fabrycky, {A'Hearn}, Wellnitz, Barry, Kuchner, Livengood, Hewagama,
  Sunshine, Hampton, Lisse, Seager, \& Veverka}]{BaChCh10}
Ballard, S., {et~al.} 2010, ApJ, 716, 1047

\bibitem[{Batalha {et~al.}(2010{\natexlab{a}})Batalha, Rowe, Gilliland,
  Jenkins, Caldwell, Borucki, Koch, Lissauer, Dunham, Gautier, Howell, Latham,
  Marcy, \& Prsa}]{BaRoGi10}
Batalha, N.~M., {et~al.} 2010{\natexlab{a}}, ApJ, 713, L103

\bibitem[{Batalha {et~al.}(2010{\natexlab{b}})Batalha, Borucki, Koch, Bryson,
  Haas, Brown, Caldwell, Hall, Gilliland, Latham, Meibom, \& Monet}]{BaBoKo10}
---. 2010{\natexlab{b}}, ApJ, 713, L109

\bibitem[{Batalha {et~al.}(2011)Batalha, Borucki, Bryson, Buchhave, Caldwell,
  Christensen-Dalsgaard, Ciardi, Dunham, Fressin, Gautier, Gilliland, Haas,
  Howell, Jenkins, Kjeldsen, Koch, Latham, Lissauer, Marcy, Rowe, Sasselov,
  Seager, Steffen, Torres, Basri, Brown, Charbonneau, Christiansen, Clarke,
  Cochran, Dupree, Fabrycky, Fischer, Ford, Fortney, Girouard, Holman, Johnson,
  Isaacson, Klaus, Machalek, Moorehead, Morehead, Ragozzine, Tenenbaum,
  Twicken, Quinn, VanCleve, Walkowicz, Welsh, Devore, \& Gould}]{BaBoBr11}
---. 2011, ApJ, 729, 27

\bibitem[{Beatty \& Seager(2010)}]{BeSe10}
Beatty, T.~G., \& Seager, S. 2010, ApJ, 712, 1433

\bibitem[{Beerer {et~al.}(2011)Beerer, Knutson, Burrows, Fortney, Agol,
  Charbonneau, Cowan, Deming, Desert, Langton, Laughlin, Lewis, \&
  Showman}]{BeKnBu11}
Beerer, I.~M., {et~al.} 2011, ApJ, 727, 23

\bibitem[{Borucki {et~al.}(2010{\natexlab{a}})Borucki, Koch, Brown, Basri,
  Batalha, Caldwell, Cochran, Dunham, Gautier, Geary, Gilliland, Howell,
  Jenkins, Latham, Lissauer, Marcy, Monet, Rowe, \& Sasselov}]{BoKoBr10}
Borucki, W.~J., {et~al.} 2010{\natexlab{a}}, ApJ, 713, L126

\bibitem[{Borucki {et~al.}(2010{\natexlab{b}})Borucki, Koch, Basri, Batalha,
  Brown, Caldwell, Caldwell, J.~{Christensen-Dalsgaard}, Cochran, DeVore,
  Dunham, Dupree, Gautier, Geary, Gilliland, Gould, Howell, Jenkins, Kondo,
  Latham, Marcy, Meibom, Kjeldsen, Lissauer, Monet, Morrison, Sasselov, Tarter,
  Boss, Brownlee, Owen, Buzasi, Charbonneau, Doyle, Fortney, Ford, Holman,
  S.Seager, Steffen, Welsh, Rowe, Anderson, Buchhave, Ciardi, Walkowicz,
  Sherry, Horch, Isaacson, Everett, Fischer, Torres, Johnson, Endl, MacQueen,
  Dotson, Haas, Kolodziejczak, {Van Cleve}and H.~Chandrasekaran, Twicken,
  Quintana, Clarke, Allen, Li, Wu, Tenenbaum, Verner, Bruhweiler, Barnes, \&
  Prsa}]{BoKoBa10}
---. 2010{\natexlab{b}}, Science, 327, 977

\bibitem[{Borucki {et~al.}(2011{\natexlab{a}})Borucki, Koch, Basri, Batalha,
  Boss, Brown, Caldwell, Christensen-Dalsgaard, Cochran, DeVore, Dunham,
  Dupree, Gautier, Geary, Gilliland, Gould, Howell, Jenkins, Kjeldsen, Latham,
  Lissauer, Marcy, Monet, Sasselov, Tarter, Charbonneau, Doyle, Ford, Fortney,
  Holman, Seager, Steffen, Welsh, Allen, Bryson, Buchhave, Chandrasekaran,
  Christiansen, Ciardi, Clarke, Dotson, Endl, Fischer, Fressin, Haas, Horch,
  Howard, Isaacson, Kolodziejczak, Li, MacQueen, Meibom, an, Quintana, Rowe,
  Sherry, Tenenbaum, Torres, Twicken, {Van Cleve}, Walkowicz, \&
  Wu}]{BoKoBa11a}
---. 2011{\natexlab{a}}, ApJ, 728, 117

\bibitem[{Borucki {et~al.}(2011{\natexlab{b}})Borucki, Koch, Basri, Batalha,
  Brown, Bryson, Caldwell, {Christensen-Dalsgaard}, Cochran, DeVore, Dunham,
  {Gautier III}, Geary, {Gilliland}, {Gould}, {Howell}, {Jenkins}, {Latham},
  {Lissauer}, {Marcy}, {Rowe}, {Sasselov}, {Boss}, {Charbonneau}, {Ciardi},
  {Doyle}, {Dupree}, {Ford}, {Fortney}, {Holman}, {Seager}, {Steffen},
  {Tarter}, {Welsh}, {Allen}, {Buchhave}, {Christiansen}, {Clarke},
  {D{\'e}sert}, {Endl}, {Fabrycky}, {Fressin}, {Haas}, {Horch}, {Howard},
  {Isaacson}, {Kjeldsen}, {Kolodziejczak}, {Kulesa}, {Li}, {Machalek},
  {McCarthy}, {MacQueen}, {Meibom}, {Miquel}, {Prsa}, {Quinn}, {Quintana},
  {Ragozzine}, {Sherry}, {Shporer}, {Tenenbaum}, {Torres}, {Twicken}, {Van
  Cleve}, \& {Walkowicz}}]{BoKoBa11b}
---. 2011{\natexlab{b}}, ApJ, in press

\bibitem[{{Brown} {et~al.}(2011){Brown}, {Latham}, {Everett}, \&
  {Esquerdo}}]{BrLaEv11}
{Brown}, T.~M., {Latham}, D.~W., {Everett}, M.~E., \& {Esquerdo}, G.~A. 2011,
  AJ, submitted, arXiv:1102.0342

\bibitem[{Brugamyer {et~al.}(2011)Brugamyer, {Dodson-Robinson}, Cochran, \&
  Sneden}]{BrDRCo11}
Brugamyer, E., {Dodson-Robinson}, S.~E., Cochran, W.~D., \& Sneden, C. 2011,
  ApJ, in press, arXiv:1106.5509

\bibitem[{Bryson {et~al.}(2010)Bryson, Tenenbaum, Jenkins, Chandrasekaran,
  Klaus, Caldwell, Gilliland, Haas, Dotson, Koch, , \& Borucki}]{BrTeJe10}
Bryson, S.~T., {et~al.} 2010, ApJ, 713, L97

\bibitem[{Butler {et~al.}(1996)Butler, Marcy, Williams, McCarthy, Dosanjh, \&
  Vogt}]{BuMaWi96}
Butler, R.~P., Marcy, G.~W., Williams, E., McCarthy, C., Dosanjh, P., \& Vogt,
  S.~S. 1996, PASP, 108, 500

\bibitem[{Caldwell {et~al.}(2010)Caldwell, Kolodziejczak, {Van Cleve}, Jenkins,
  Gazis, Argabright, Bachtell, Dunham, Geary, Gilliland, Chandrasekaran, Li,
  Tenenbaum, Wu, J.Borucki, Bryson, Dotson, Haas, \& G.Koch}]{CaKoVC10}
Caldwell, D.~A., {et~al.} 2010, ApJ, 713, L92

\bibitem[{Charbonneau {et~al.}(2005)Charbonneau, Allen, Megeath, Torres,
  Alonso, Brown, Gilliland, Latham, Mandushev, {O'Donovan}, \&
  Sozzetti}]{ChAlMe05}
Charbonneau, D., {et~al.} 2005, ApJ, 626, 523

\bibitem[{Charbonneau {et~al.}(2009)Charbonneau, Berta, Irwin, Burke, Nutzman,
  Buchhave, Lovis, Bonfils, Latham, Udry, {Murray-Clay}, Holman, Falco, Winn,
  Queloz, Pepe, Mayor, Delfosse, \& Forveille}]{ChBeIr09}
---. 2009, Nature, 462, 891

\bibitem[{Cochran {et~al.}(2002)Cochran, Hatzes, \& Paulson}]{CoHaPa02}
Cochran, W.~D., Hatzes, A.~P., \& Paulson, D.~B. 2002, AJ, 124, 565

\bibitem[{Deming {et~al.}(2011)Deming, Knutson, Agol, Desert, Burrows, Fortney,
  Charbonneau, Cowan, Laughlin, Langton, Showman, \& Lewis}]{DeKnAg11}
Deming, D., {et~al.} 2011, ApJ, 726, 95

\bibitem[{D{\'e}sert {et~al.}(2009)D{\'e}sert, {Lecavelier des Etangs},
  H{\'e}brard, Sing, Ehrenreich, Ferlet, \& {Vidal-Madjar}}]{DeEtHe09}
D{\'e}sert, J.-M., {Lecavelier des Etangs}, A., H{\'e}brard, G., Sing, D.~K.,
  Ehrenreich, D., Ferlet, R., \& {Vidal-Madjar}, A. 2009, ApJ, 699, 478

\bibitem[{D{\'e}sert {et~al.}(2011{\natexlab{a}})D{\'e}sert, Charbonneau,
  Fortney, Madhusudhan, Knutson, Fressin, Deming, Borucki, Brown, Caldwell,
  Ford, Gilliland, Latham, Marcy, Seager, \& {the Kepler Science
  Team}}]{DeChFo11}
D{\'e}sert, J.-M., {et~al.} 2011{\natexlab{a}}, ApJ, submitted, arXiv:1102.0555

\bibitem[{D{\'e}sert {et~al.}(2011{\natexlab{b}})D{\'e}sert, Sing,
  {Vidal-Madjar}, H{\'e}brard, Ehrenreich, {Lecavelier Des Etangs}, Parmentier,
  Ferlet, \& Henry}]{DeSiVM11}
---. 2011{\natexlab{b}}, A\&A, 526, A12

\bibitem[{Dunham {et~al.}(2010)Dunham, Borucki, Koch, Batalha, Buchhave, Brown,
  Caldwell, Cochran, Endl, Fischer, {F{\H u}r{\'e}sz}, Gautier, Geary,
  Gilliland, Gould, Howell, Jenkins, Kjeldsen, Latham, Lissauer, Marcy, Meibom,
  Monet, Rowe, \& Sasselov}]{DuBoKo10}
Dunham, E.~W., {et~al.} 2010, ApJ, 713, L136

\bibitem[{Eastman {et~al.}(2010)Eastman, Siverd, \& Gaudi}]{EaSiGa10}
Eastman, J., Siverd, R., \& Gaudi, B.~S. 2010, PASP, 122, 935

\bibitem[{Fabrycky(2010)}]{Fa10}
Fabrycky, D.~C. 2010, in Exoplanets, ed. {Seager, S.} (University of Arizona
  Press), 217--238

\bibitem[{{Fazio} {et~al.}(2004){Fazio}, {Hora}, {Allen}, {Ashby}, {Barmby},
  {Deutsch}, {Huang}, {Kleiner}, {Marengo}, {Megeath}, {Melnick}, {Pahre},
  {Patten}, {Polizotti}, {Smith}, {Taylor}, {Wang}, {Willner}, {Hoffmann},
  {Pipher}, {Forrest}, {McMurty}, {McCreight}, {McKelvey}, {McMurray}, {Koch},
  {Moseley}, {Arendt}, {Mentzell}, {Marx}, {Losch}, {Mayman}, {Eichhorn},
  {Krebs}, {Jhabvala}, {Gezari}, {Fixsen}, {Flores}, {Shakoorzadeh}, {Jungo},
  {Hakun}, {Workman}, {Karpati}, {Kichak}, {Whitley}, {Mann}, {Tollestrup},
  {Eisenhardt}, {Stern}, {Gorjian}, {Bhattacharya}, {Carey}, {Nelson},
  {Glaccum}, {Lacy}, {Lowrance}, {Laine}, {Reach}, {Stauffer}, {Surace},
  {Wilson}, {Wright}, {Hoffman}, {Domingo}, \& {Cohen}}]{FaHaAl04}
{Fazio}, G.~G., {et~al.} 2004, ApJS, 154, 10

\bibitem[{Fortney {et~al.}(2007)Fortney, Marley, \& W.Barnes}]{FoMaBa07}
Fortney, J.~J., Marley, M.~S., \& W.Barnes, J. 2007, ApJ, 659, 1661

\bibitem[{Fortney \& Nettelmann(2010)}]{FoNe10}
Fortney, J.~J., \& Nettelmann, N. 2010, Space Sci. Rev., 152, 423

\bibitem[{Fressin {et~al.}(2011)Fressin, Torres, D{\'e}sert, Charbonneau,
  Batalha, Fortney, Rowe, Allen, Borucki, Brown, Bryson, Ciardi, Cochran,
  Deming, Dunham, Fabrycky, {Gautier III}, Gilliland, Henze, Holman, Howell,
  Jenkins, Kamal, Kinemuchi, Knutson, Koch, Latham, Lissauer, Marcy, Ragozzine,
  Sasselov, Still, , \& Tenenbaum}]{FrToDe11}
Fressin, F., {et~al.} 2011, ApJ, in press, arXiv:1105.4647

\bibitem[{Gilliland {et~al.}(2010)Gilliland, Jenkins, Borucki, Bryson,
  Caldwell, Clarke, Dotson, Haas, Hall, Klaus, Koch, McCauliff, Quintana,
  Twicken, \& {van Cleve}}]{GiJeBo10}
Gilliland, R.~L., {et~al.} 2010, ApJ, 713, L160

\bibitem[{Gregory(2011)}]{Gr11a}
Gregory, P.~C. 2011, MNRAS, 410, 94

\bibitem[{Halbwachs {et~al.}(2003)Halbwachs, Mayor, Udry, \& Arenou}]{HaMaUd03}
Halbwachs, J.~L., Mayor, M., Udry, S., \& Arenou, F. 2003, A\&A, 397, 159

\bibitem[{Hartman {et~al.}(2011)Hartman, Bakos, Torres, Kov{\'a}cs, Noyes,
  Latham, Howard, Fischer, Johnson, Marcy, Isaacson, Quinn, Buchhave, B{\'e}ky,
  Sasselov, Stefanik, Esquerdo, Everett, Perumpilly, L{\'a}z{\'a}r, Papp, \&
  S{\'a}ri}]{HaBaKi11}
Hartman, J.~D., {et~al.} 2011, ApJ, 728, 138

\bibitem[{Hatzes {et~al.}(2011)Hatzes, Fridlund, Nachmani, Mazeh, Valencia,
  Hebrard, Carone, Paetzold, Udry, Bouchy, Borde, Deeg, Tingley, Dvorak,
  Gandolfi, {Ferraz-Mello}, Wuchterl, Guenther, Rauer, Erikson, Cabrera,
  Csizmadia, Leger, Lammer, Weingrill, Queloz, Alonso, \& Schneider}]{HaFrNa11}
Hatzes, A.~P., {et~al.} 2011, ApJ, submitted, arXiv:1105.3372

\bibitem[{Hayward {et~al.}(2001)Hayward, Brandl, Pirger, Blacken, Gull,
  Schoenwald, \& Houck}]{HaBrPi01}
Hayward, T.~L., Brandl, B., Pirger, B., Blacken, C., Gull, G.~E., Schoenwald,
  J., \& Houck, J.~R. 2001, PASP, 113, 105

\bibitem[{Holman {et~al.}(2010)Holman, Fabrycky, Ragozzine, Ford, Steffen,
  Welsh, Lissauer, Latham, Marcy, Walkowicz, Batalha, Jenkins, Rowe, Cochran,
  Fressin, Torres, Buchhave, Sasselov, Borucki, Koch, Basri, Brown, Caldwell,
  Charbonneau, Dunham, Gautier, Geary, Gilliland, Haas, Howell, Ciardi, Endl,
  Fischer, {F{\"u}r{\'e}sz}, Hartman, Isaacson, Johnson, MacQueen, Moorhead,
  Morehead, \& A.Orosz}]{HoFaRa10}
Holman, M.~J., {et~al.} 2010, Science, 330, 51

\bibitem[{Horch {et~al.}(2011)Horch, Gomez, Sherry, Howell, Ciardi, Anderson,
  \& {van Altena}}]{HoGoSh11}
Horch, E.~P., Gomez, S.~C., Sherry, W.~H., Howell, S.~B., Ciardi, D.~R.,
  Anderson, L.~M., \& {van Altena}, W.~F. 2011, AJ, 141, 45

\bibitem[{Howard {et~al.}(2010)Howard, Marcy, Johnson, Fischer, Wright,
  Isaacson, Valenti, Anderson, Lin, \& Ida}]{HoMaJo10}
Howard, A.~W., {et~al.} 2010, Science, 330, 653

\bibitem[{Howell {et~al.}(2011)Howell, Everett, Sherry, Horch, \&
  Ciardi}]{HoEvSh11}
Howell, S.~B., Everett, M.~E., Sherry, W., Horch, E., \& Ciardi, D.~R. 2011,
  AJ, 42, 19

\bibitem[{Jackson {et~al.}(2008)Jackson, Greenberg, \& Barnes}]{JaGrBa08}
Jackson, B., Greenberg, R., \& Barnes, R. 2008, ApJ, 681, 1631

\bibitem[{Jenkins {et~al.}(2010{\natexlab{a}})Jenkins, Borucki, Koch, Marcy,
  Cochran, Basri, Batalha, Buchhave, Brown, Caldwell, Dunham, Endl, Fischer,
  {Gautier III}, Geary, Gilliland, Howell, Isaacson, Johnson, {Latham},
  {Lissauer}, {Monet}, {Rowe}, {Sasselov}, {Welsh}, {Howard}, {MacQueen},
  {Chandrasekaran}, {Twicken}, {Bryson}, {Quintana}, {Clarke}, {Li}, {Allen},
  {Tenenbaum}, {Wu}, {Meibom}, {Klaus}, {Middour}, {Cote}, {McCauliff},
  {Girouard}, {Gunter}, {Wohler}, {Hall}, {Ibrahim}, {Kamal Uddin}, {Wu},
  {Bhavsar}, {Van Cleve}, {Pletcher}, {Dotson}, \& {Haas}}]{JeBoKo10}
Jenkins, J.~M., {et~al.} 2010{\natexlab{a}}, ApJ, 724, 1108

\bibitem[{Jenkins {et~al.}(2010{\natexlab{b}})Jenkins, Caldwell,
  Chandrasekaran, Twicken, Bryson, Quintana, Clarke, Li, Allen, Tenenbaum, Wu,
  Klaus, {Van Cleve}, Dotson, Haas, Gilliland, Koch, \& Borucki}]{JeCaCh10a}
---. 2010{\natexlab{b}}, ApJ, 713, L120

\bibitem[{Jenkins {et~al.}(2010{\natexlab{c}})Jenkins, Caldwell,
  Chandrasekaran, Twicken, Bryson, Quintana, Clarke, Li, Allen, Tenenbaum, Wu,
  Klaus, Middour, Cote, McCauliff, Girouard, Gunter, Wohler, Sommers, Hall,
  Uddin, Wu, Bhavsar, {Van Cleve}, Pletcher, Dotson, Haas, Gilliland, Koch, \&
  Borucki}]{JeCaCh10b}
---. 2010{\natexlab{c}}, ApJ, 713, L87

\bibitem[{Jenkins {et~al.}(2010{\natexlab{d}})Jenkins, Chandrasekaran,
  McCauliff, Caldwell, Tenenbaum, Li, Klaus, Cote, \& Middour}]{JeCChMC10}
---. 2010{\natexlab{d}}, Proc. Soc. Photo-opt. Inst. Eng., 7740, D

\bibitem[{Johnson {et~al.}(2009)Johnson, Winn, Albrecht, Howard, Marcy, \&
  Gazak}]{JoWiAl09}
Johnson, J.~A., Winn, J.~N., Albrecht, S., Howard, A.~W., Marcy, G.~W., \&
  Gazak, J.~Z. 2009, PASP, 121, 1104

\bibitem[{Knutson {et~al.}(2008)Knutson, Charbonneau, Allen, Burrows, \&
  Megeath}]{KnChAl08}
Knutson, H.~A., Charbonneau, D., Allen, L.~E., Burrows, A., \& Megeath, S.~T.
  2008, ApJ, 673, 526

\bibitem[{Koch {et~al.}(2010{\natexlab{a}})Koch, Borucki, Rowe, Batalha, Brown,
  Caldwell, Caldwell, Cochran, DeVore, Dunham, Dupree, Gautier, Geary,
  Gilliland, Howell, Jenkins, Latham, Lissauer, Marcy, Morrison, \&
  Tarter}]{KoBoRo10}
Koch, D.~G., {et~al.} 2010{\natexlab{a}}, ApJ, 713, L131

\bibitem[{Koch {et~al.}(2010{\natexlab{b}})Koch, Borucki, Basri, Batalha,
  Brown, Caldwell, Christensen-Dalsgaard, Cochran, DeVore, Dunham, Gautier,
  Geary, Gilliland, Gould, Jenkins, Kondo, Latham, Lissauer, Marcy, Monet,
  Sasselov, Boss, Brownlee, Caldwell, Dupree, Howell, Kjeldsen, Meibom,
  Morrison, Owen, Reitsema, Tarter, Bryson, Dotson, Gazis, Haas, Kolodziejczak,
  Rowe, Cleve, Allen, Chandrasekaran, Clarke, Li, Quintana, Tenenbaum, Twicken,
  \& Wu}]{KoBoBa10}
---. 2010{\natexlab{b}}, ApJ, 713, L79

\bibitem[{Latham {et~al.}(2010)Latham, Borucki, Koch, Brown, Buchhave, Basri,
  Batalha, Caldwell, Cochran, Dunham, {F{\H u}r{\'e}sz}, Gautier, Geary,
  Gilliland, Howell, Jenkins, Lissauer, Marcy, Monet, Rowe, \&
  Sasselov}]{LaBoKo10}
Latham, D.~W., {et~al.} 2010, ApJ, 713, L140

\bibitem[{{Lissauer} {et~al.}(2011{\natexlab{a}}){Lissauer}, {Fabrycky},
  {Ford}, {Borucki}, {Fressin}, {Marcy}, {Orosz}, {Rowe}, {Torres}, {Welsh},
  {Batalha}, {Bryson}, {Buchhave}, {Caldwell}, {Carter}, {Charbonneau},
  {Christiansen}, {Cochran}, {Desert}, {Dunham}, {Fanelli}, {Fortney},
  {Gautier}, {Geary}, {Gilliland}, {Haas}, {Hall}, {Holman}, {Koch}, {Latham},
  {Lopez}, {McCauliff}, {Miller}, {Morehead}, {Quintana}, {Ragozzine},
  {Sasselov}, {Short}, \& {Steffen}}]{LiFaFo11}
{Lissauer}, J.~J., {et~al.} 2011{\natexlab{a}}, Nature, 470, 53

\bibitem[{{Lissauer} {et~al.}(2011{\natexlab{b}}){Lissauer}, {Ragozzine},
  {Fabrycky}, {Steffen}, {Ford}, {Jenkins}, {Shporer}, {Holman}, {Rowe},
  {Quintana}, {Batalha}, {Borucki}, {Bryson}, {Caldwell}, {Carter}, {Ciardi},
  {Dunham}, {Fortney}, {Gautier}, {Howell}, {Koch}, {Latham}, {Marcy},
  {Morehead}, \& {Sasselov}}]{LiRaFa11}
---. 2011{\natexlab{b}}, ApJ, in press, 1102.0543

\bibitem[{Mandel \& Agol(2002)}]{MaAg02}
Mandel, K., \& Agol, E. 2002, ApJ, 580, L171

\bibitem[{Marcus {et~al.}(2010)Marcus, Sasselov, Hernquist, \&
  Stewart}]{MaSaHe10}
Marcus, R.~A., Sasselov, D., Hernquist, L., \& Stewart, S.~T. 2010, ApJ, 712,
  L73

\bibitem[{Marcy {et~al.}(2008)Marcy, Butler, Vogt, Fischer, Wright, Johnson,
  Tinney, Jones, Carter, Bailey, O'Toole, \& S.~Upadhyay}]{MaBuVo08}
Marcy, G.~W., {et~al.} 2008, Physica Scripta, T130, 14001

\bibitem[{Markwardt(2009)}]{Ma09}
Markwardt, C.~B. 2009, in Astronomical Society of the Pacific Conference
  Series, Vol. 411, Astronomical Data Analysis Software and Systems XVIII, ed.
  {D.~A.~Bohlender, D.~Durand, \& P.~Dowler}, 251

\bibitem[{Mazeh(2008)}]{Ma08}
Mazeh, T. 2008, in EAS Publications Series, ed. {M.-J.~Goupil \& J.-P.~Zahn},
  Vol.~29, 1

\bibitem[{Miller \& Fortney(2011)}]{MiFo11}
Miller, N., \& Fortney, J.~J. 2011, ApJ, in press, arXiv:1105.0024

\bibitem[{Miller {et~al.}(2009)Miller, Fortney, \& Jackson}]{MiFoJa09}
Miller, N., Fortney, J.~J., \& Jackson, B. 2009, ApJ, 702, 1413

\bibitem[{{Miralda-Escud{\'e}}(2002)}]{ME02}
{Miralda-Escud{\'e}}, J. 2002, ApJ, 564, 1019

\bibitem[{{Morton} \& {Johnson}(2011)}]{MoJo11}
{Morton}, T.~D., \& {Johnson}, J.~A. 2011, ApJ, in press, arXiv:1101.5630

\bibitem[{Pont {et~al.}(2006)Pont, Zucker, \& Queloz}]{PoZuQu06}
Pont, F., Zucker, S., \& Queloz, D. 2006, MNRAS, 373, 231

\bibitem[{Queloz {et~al.}(2009)Queloz, Bouchy, Moutou, Hatzes, H{\'e}brard,
  Alonso, Auvergne, Baglin, Barbieri, Barge, Benz, Bord{\'e}, Deeg, Deleuil,
  Dvorak, Erikson, {Ferraz Mello}, Fridlund, Gandolfi, Gillon, Guenther, Jorda,
  Hartmann, Lammer, L{\'e}ger, Llebaria, Lovis, Magain, Mayor, Mazeh, Ollivier,
  P{\"a}tzold, Pepe, Rauer, Rouan, Schneider, Segransan, Udry, \&
  Wuchterl}]{QuBoMo09}
Queloz, D., {et~al.} 2009, A\&A, 506, 303

\bibitem[{{Raghavan} {et~al.}(2010){Raghavan}, {McAlister}, {Henry}, {Latham},
  {Marcy}, {Mason}, {Gies}, {White}, \& {ten Brummelaar}}]{RaMAHe10}
{Raghavan}, D., {et~al.} 2010, ApJS, 190, 1

\bibitem[{Ragozzine \& Holman(2010)}]{RaHo10}
Ragozzine, D., \& Holman, M.~J. 2010, ApJ, submitted, arXiv:1006.3727

\bibitem[{{Robin} {et~al.}(2003){Robin}, {Reyl{\'e}}, {Derri{\`e}re}, \&
  {Picaud}}]{RoReDe03}
{Robin}, A.~C., {Reyl{\'e}}, C., {Derri{\`e}re}, S., \& {Picaud}, S. 2003,
  A\&A, 409, 523

\bibitem[{Rogers {et~al.}(2011)Rogers, Bodenheimer, Lissauer, \&
  Seager}]{RoBoLi11}
Rogers, L.~A., Bodenheimer, P., Lissauer, J.~J., \& Seager, S. 2011, ApJ, in
  press, arXiv:1106.2807

\bibitem[{Seager {et~al.}(2007)Seager, Kuchner, {Hier-Majumder}, \&
  Militzer}]{SeKuHM07}
Seager, S., Kuchner, M., {Hier-Majumder}, C.~A., \& Militzer, B. 2007, ApJ,
  669, 1279

\bibitem[{Sneden(1973)}]{Sn73}
Sneden, C.~A. 1973, PhD thesis, {The University of Texas at Austin}

\bibitem[{Steffen {et~al.}(2010)Steffen, Batalha, Borucki, Buchhave, Caldwell,
  Cochran, Endl, Fabrycky, Fressin, Ford, Fortney, Haas, Holman, Howell,
  Isaacson, Jenkins, Koch, Latham, Lissauer, Moorhead, Morehead, Marcy,
  MacQueen, Quinn, Ragozzine, Rowe, Sasselov, Seager, Torres, \&
  Welsh}]{StBaBo10}
Steffen, J.~H., {et~al.} 2010, ApJ, 725, 1226

\bibitem[{{Torres} {et~al.}(2004){Torres}, {Konacki}, {Sasselov}, \&
  {Jha}}]{ToKoSa04}
{Torres}, G., {Konacki}, M., {Sasselov}, D.~D., \& {Jha}, S. 2004, ApJ, 614,
  979

\bibitem[{{Torres} {et~al.}(2011){Torres}, {Fressin}, {Batalha}, {Borucki},
  {Brown}, {Bryson}, {Buchhave}, {Charbonneau}, {Ciardi}, {Dunham}, {Fabrycky},
  {Ford}, {Gautier}, {Gilliland}, {Holman}, {Howell}, {Isaacson}, {Jenkins},
  {Koch}, {Latham}, {Lissauer}, {Marcy}, {Monet}, {Prsa}, {Quinn}, {Ragozzine},
  {Rowe}, {Sasselov}, {Steffen}, \& {Welsh}}]{ToFrBa11}
{Torres}, G., {et~al.} 2011, ApJ, 727, 24

\bibitem[{Troy {et~al.}(2000)Troy, Dekany, Brack, Oppenheimer, Bloemhof, Trinh,
  Dekens, Shi, Hayward, \& Brandl}]{TrDeBr00}
Troy, M., {et~al.} 2000, Proc. Soc. Photo-opt. Inst. Eng., 4007, 31

\bibitem[{Valencia {et~al.}(2006)Valencia, O'Connell, \& Sasselov}]{VaOCSa06}
Valencia, D., O'Connell, R.~J., \& Sasselov, D. 2006, Icarus, 181, 545

\bibitem[{Valenti \& Fischer(2005)}]{VaFi05}
Valenti, J.~A., \& Fischer, D.~A. 2005, ApJS, 159, 141

\bibitem[{Valenti \& Piskunov(1996)}]{VaPi96}
Valenti, J.~A., \& Piskunov, N. 1996, A\&AS, 118, 595

\bibitem[{Vogt {et~al.}(1994)Vogt, Allen, Bigelow, Bresee, Brown, Cantrall,
  Conrad, Couture, Delaney, Epps, Hilyard, Hilyard, Horn, Jern, Kanto, Keane,
  Kibrick, Lewis, Osborne, Pardeilhan, Pfister, Ricketts, Robinson, Stover,
  Tucker, Ward, \& Wei}]{VoAlBi94}
Vogt, S.~S., {et~al.} 1994, Proc. Soc. Photo-opt. Inst. Eng., 2198, 362

\bibitem[{Werner {et~al.}(2004)Werner, Roellig, Low, Rieke, Rieke, Hoffmann,
  Young, Houck, Brandl, Fazio, Hora, Gehrz, Helou, Soifer, Stauffer, Keene,
  Eisenhardt, Gallagher, Gautier, Irace, Lawrence, Simmons, Cleve, Jura,
  Wright, \& Cruikshank}]{WeRoLo04}
Werner, M.~W., {et~al.} 2004, ApJS, 154, 1

\bibitem[{Winn {et~al.}(2011)Winn, Matthews, Dawson, Fabrycky, Holman,
  Kallinger, Kuschnig, Sasselov, Dragomir, Guenther, Moffat, Rowe, Rucinski, \&
  Weiss}]{WiMaDa11}
Winn, J.~N., {et~al.} 2011, ApJ, in press, arXiv:1104.5230

\bibitem[{Wu {et~al.}(2010)Wu, Twicken, Tenenbaum, Clarke, Li, Quintana, Allen,
  Chandrasekaran, Jenkins, Caldwell, Wohler, Girouard, McCauliff, Cote, \&
  Klaus}]{WuTwTe10}
Wu, H., {et~al.} 2010, Proc. Soc. Photo-opt. Inst. Eng., 7740, 19

\bibitem[{Yi {et~al.}(2001)Yi, Demarque, Kim, Lee, Ree, Lejeune, \&
  Barnes}]{YiDeKi01}
Yi, S., Demarque, P., Kim, Y.-C., Lee, Y.-W., Ree, C.~H., Lejeune, T., \&
  Barnes, S. 2001, ApJS, 136, 417

\bibitem[{{Zechmeister} \& {K{\"u}rster}(2009)}]{ZeKu09}
{Zechmeister}, M., \& {K{\"u}rster}, M. 2009, A\&A, 496, 577

\end{thebibliography}
\end{document}